%% file: paper_arxiv.tex
\documentclass[preprint,12pt]{elsarticle}
\usepackage{graphics}
\usepackage{epsfig}
\usepackage{amssymb}
\usepackage{multirow}

\input{format}

\journal{Elsevier}

\begin{document}

\begin{frontmatter}

\title{Computational homogenization of non-stationary transport processes in masonry
structures}

\author[ctu]{Jan S\'{y}kora}
\ead{jan.sykora.1@fsv.cvut.cz}
\author[ctu]{Tom\'{a}\v{s} Krej\v{c}\'{i}}
\ead{krejci@cml.fsv.cvut.cz}
\author[ctu]{Jaroslav Kruis}
\ead{jk@cml.fsv.cvut.cz}
\author[ctu]{Michal \v{S}ejnoha\corref{auth}}
\ead{sejnom@fsv.cvut.cz}
\cortext[auth]{Corresponding author. Tel.:~+420-2-2435-4494;
fax~+420-2-2431-0775}
\address[ctu]{Department of Mechanics, Faculty of Civil Engineering,
  Czech Technical University in Prague, Th\'{a}kurova 7, 166 29 Prague
  6, Czech Republic}

\begin{abstract}
A fully coupled transient heat and moisture transport in a masonry
structure is examined in this paper. Supported by several successful
applications in civil engineering the nonlinear diffusion model
proposed by K\"{u}nzel~\cite{Kunzel:IJHMT:1997} is adopted in the
present study. A strong material heterogeneity together with
a significant dependence of the model parameters on initial conditions as
well as the gradients of heat and moisture fields vindicates the use of
a hierarchical modeling strategy to solve the problem of this kind.
Attention is limited to the classical first order homogenization in a
spatial domain developed here in the framework of a two step
(meso-macro) multi-scale computational scheme (FE$^2$
problem). Several illustrative examples are presented to investigate
the influence of transient flow at the level of constituents
(meso-scale) on the macroscopic response including the effect of
macro-scale boundary conditions. A two-dimensional section of Charles
Bridge subjected to actual climatic conditions is analyzed next to
confirm the suitability of algorithmic format of FE$^2$ scheme for
the parallel computing.
\end{abstract}

\end{frontmatter}

\section{Introduction}\label{sec:Introduction}
Historical masonry structures all over the world enjoy constant
attention by many entities including technical audience and public
authorities. When subject to restoration these structures naturally
invite complex analyses combining both experimental
work~\cite{Prikryl:2010:IG} and numerical
simulations~\cite{Novak:ES:2007}. Even after closing the scheduled
reconstruction steps a continuation of in-situ monitoring is now
becoming almost standard allowing the engineers not only to evaluate
the current state of the structure but also to improve the predictive
capability of theoretical models being supported by up to date
material data and instantaneous measurements of the state of stress
and deformation. A particularly vivid example of this line of inquiry
is the Charles Bridge information system~\cite{CHBIS} integrating
detailed geometrical description of the bridge including changes
resulting from previous reconstructions, historical and contemporary
materials forming the bridge as well as novel materials and
technologies used to improve its durability and serviceable life.  The
available results of long-term measurements of temperature and
moisture fields, time-dependent displacement data, complemented by
advanced methods of computational analysis then open the way to
incorporate multi-physics, multi-scale, time-dependent and
three-dimensional aspects of the problem to create a realistic
computational model of such a complex structure.

These issues have been supported by our recent
study~\cite{Sykora:2011:AMC} devoted to the homogenization of masonry
walls with emphases on the effect of imperfect hydraulic contact. On
the one hand, it has been demonstrated that the introduction of
interface transition zone at the mesostructural level, considerably
complicating the computational matter, is essentially negligible for
the prediction of effective properties. On the other hand, an
accompanied parametric study has shown a substantial dependence of the
homogenized properties on both the initial and loading conditions.
Keeping the format of a simplified uncoupled multi-scale scheme with
several successful applications particularly in civil
engineering~\cite[to cite a
  few]{Novak:ES:2007,Valenta:IJMCE:10,Sejnoha:CC:2011} would require
performing the homogenization analysis for various values of water
content and certain referenced initial values of temperature and
relative humidity to construct the homogenized macro-scale retention
curves. Utilizing these curves in an independent macroscopic study
would considerably increase the computational efficiency since
avoiding a time consuming down scaling for the parameters update at
every macroscopic time step. Unfortunately, an observed strong
dependence of the effective properties on the applied macroscopic
gradients may lead to an enormous database of these response functions
essentially loosing the advantage over a full-fledged coupled
multi-scale framework. This issue together with the possibility of
including the mesostructural morphology and mesostructural material
behavior in the macro-level, where typical structures are analyzed,
without the need for assigning the fine-scale details to the entire
structure thus motivated the developments in this paper towards an
iterative FE$^2$ algorithm much similar to that presented
in~\cite{Larsson:2010:IJMNE}.

Owing to the finite size of the representative volume element on
meso-scale we adopt a variationally consistent homogenization
developed in~\cite{Larsson:2010:IJMNE} to reflect a certain size
dependency of distributions of macroscopic fields due to higher order
terms appearing on the left hand side of macroscopic balance equations
when a non-linear transient flow is assumed on both the macro and
meso-scale.  A short numerical study of this topic is presented in
Section~\ref{subsec:InfTRANS} preceded by the theoretical formulation
in Section~\ref{sec:Homogenization}. Section~\ref{subsec:InfFEM} then
investigates the influence of macroscopic finite element mesh and
application of the associated loading conditions in FE$^2$ scheme on
the accuracy of evolution of macroscopic moisture and temperature
fields. The obtained results are then utilized in
Section~\ref{subsec:Parallelization} devoted to the parallel format of
the underlying multi-scale analysis performed for a two-dimensional
section of Charles Bridge. Summary of the essential findings is
available in Section~\ref{sec:Conclusions}. To keep the paper
self-contained a short overview of the adopted constitutive model is
provided in Section~\ref{sec:Matmodel}.

In the following text, $\vek{a}$ and $\tensf{A}$ denote a vector
and a symmetric second-order tensor, respectively. The symbol
$\vek{\nabla}=\left\{\displaystyle{{\partial}/{\partial{x}}},
\displaystyle{{\partial}/{\partial{y}}},\displaystyle{{\partial}/{\partial{z}}}\right\}^{\sf
T}$ stands for the gradient representation. All materials are
assumed locally isotropic.

\section{Material model}\label{sec:Matmodel}
The literature offers a manifold of material models that allow for the
description of coupled heat and moisture transport. An extensive
overview of transport models is presented in the monograph by
\v{C}ern\'{y} and Rovnan\'{\i}kov\'{a}~\cite{Cerny:2002}. Among others
  the non-linear diffusion model proposed by K\"{u}nzel holds a great
  potential for an accurate description of transport processes in
  building engineering including masonry structures and will be
  adopted here to follow up our previous works in this
  area~\cite{Sejnoha:MS:2008,Sykora:2011:AMC}.

K\"{u}nzel derived the coupled system of energy and mass balance
equations based on concepts put forward by Krischer and Kiessl, see
e.g.~\cite{Kunzel:1995,Kunzel:IJHMT:1997}. In~\cite{Krischer:1978}
Krischer identified two transport mechanisms for material moisture,
one being the vapor diffusion and the other being described as a
capillary water movement. In other words, Krischer introduced the
gradient of partial pressure in the air as the driving force for the
water vapor transport and the gradient of liquid moisture content as
the driving force for the water transport. Kiessl further extended the
diffusion model of Krischer and developed in ~\cite{Kiessl:1983} his
own original version. The unification for the description of moisture
transport in the hygroscopic $\varphi\leq 0.9$ and overhygroscopic
$\varphi > 0.9$ range ($\varphi$ is the relative humidity) was
achieved with the help of moisture potential, which brought several
advantages particularly a very simple expression for the moisture
transport across interfaces. On the other hand, the definition of
moisture potential in the overhygroscopic range was too artificial,
and Kiessl introduced it without any theoretical background,
see~\cite{Cerny:2002}.

For the description of simultaneous water and water vapor transport
K\"{u}nzel neglected the liquid water and water vapor convection
driven by gravity and total pressure as well as enthalpy changes due
to liquid flow and choose the relative humidity $\varphi$ as the only
moisture potential for both hygroscopic and overhygroscopic range. He
also divided overhygroscopic region into two sub-ranges - capillary
water region and supersaturated region, where different conditions for
water and water vapor transport are considered. In comparison with
Kiessl's or Krischer's model K\"{u}nzel's model introduces several
simplifications. Nevertheless, the proposed model describes all
substantial phenomena of the heat and moisture transport in building
materials and the predicted results comply well with the
experimentally obtained data \cite{Sykora:2011:AMC}.

Employing the classical Fick's law for the description of water vapor
diffusion, Kelvin's law to simulate the transport of liquid water and
Fourier's law to account for the flow of heat energy the K\"unzel model
integrates into the energy balance equation
\begin{equation}\label{eq:Kun01}
\frac{\mathrm{d}H}{\mathrm{d}\theta}\frac{\mathrm{d}\theta}{\mathrm{d}t}
 =  \vek{\nabla}^{\mathrm{T}}[\lambda\vek{\nabla} \theta]+h_{v}\vek{\nabla}^{\mathrm{T}}
[\delta_{p}\vek{\nabla}\{\varphi p_{\mathrm{sat}}(\theta)\}],
\end{equation}
and the conservation of mass equation
\begin{equation}\label{eq:Kun02}
\frac{\mathrm{d}w}{\mathrm{d}\varphi}\frac{\mathrm{d}
\varphi}{\mathrm{d}t}  =  \vek{\nabla}^{\mathrm{T}}
[D_{\varphi}\vek{\nabla}\varphi]+\vek{\nabla}^{\mathrm{T}}
[\delta_{p}\vek{\nabla}\{\varphi p_{\mathrm{sat}}(\theta)\}] \, ,
\end{equation}
where $H$ is the enthalpy of the moist building material, $w$ is the
water content of the building material, $\lambda$ is the thermal
conductivity, $D_{\varphi}$ is the liquid conduction coefficient,
$\delta_{p}$ is the water vapor permeability, $h_{v}$ is the
evaporation enthalpy of the water, $p_{\mathrm{sat}}$ is the water
vapor saturation pressure, $\theta$ is the temperature and $\varphi$
is the relative humidity. The material parameters, both measured and
functionally derived, that enter the above equations are summarized
for the sake of completeness in~\ref{app:A}, see
also~\cite{Kunzel:1995,Sykora:2010} for more detailed discussion on
this subject. Note that the second term on the right hand side of
Eq.~\eqref{Kun01} represents the change of enthalpy due to phase
transition being considered the only heat source or sink.

\begin{figure}[ht]
\begin{center}
\begin{tabular}{c}
\includegraphics*[width=65mm,keepaspectratio]{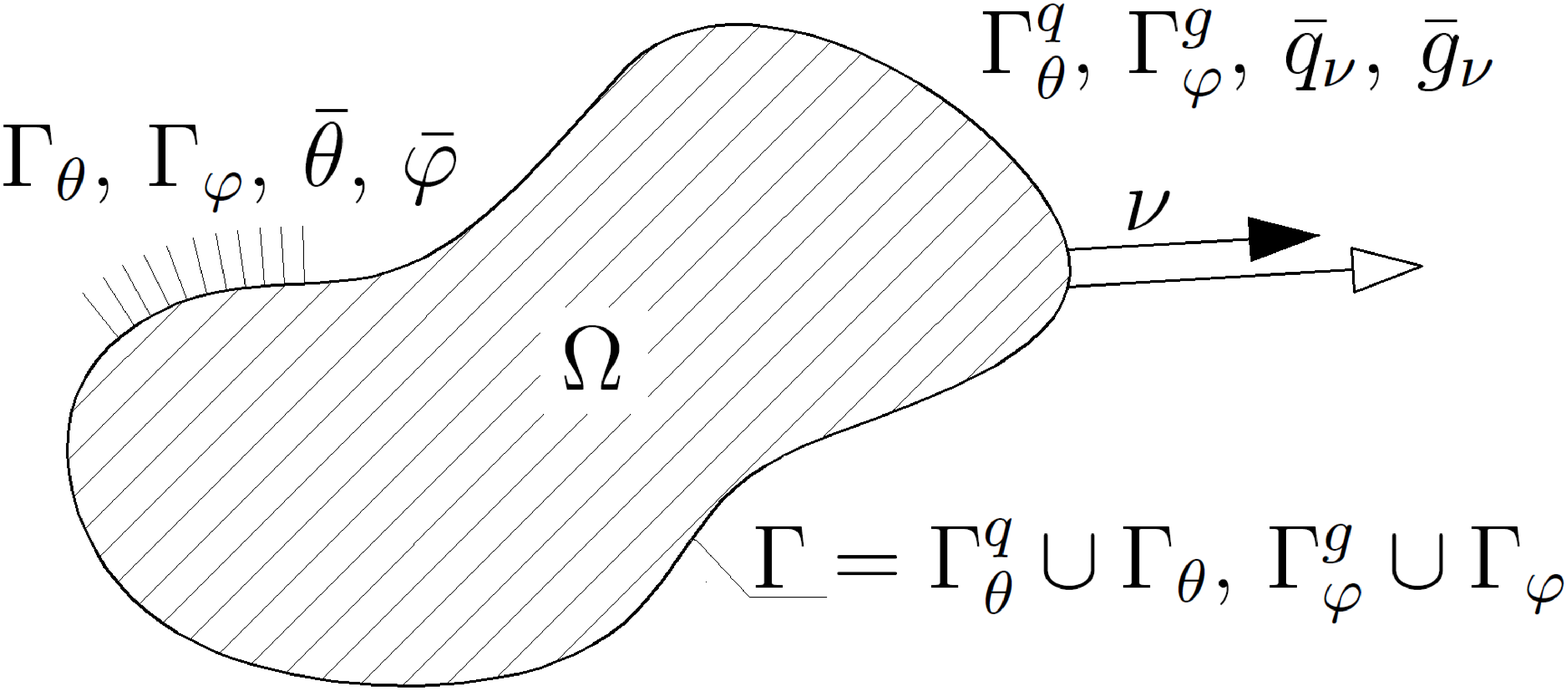}
\end{tabular}
\end{center}
\caption{A two dimensional region}
\label{fig:DDE1}
\end{figure}

Application of principal of virtual work then yields the weak form of
these two balance equations
\begin{eqnarray}
&&\int_{\Omega}\delta\theta
\frac{\mathrm{d}H}{\mathrm{d}\theta}\frac{\mathrm{d}\theta}{\mathrm{d}t}\mathrm{d}\Omega
+\int_{\Omega}\{\vek{\nabla}\delta
\theta\}^{\mathrm{\textsf{T}}}\left[\{\mat{\lambda}\vek{\nabla}
\theta\}+h_{v}
\left\{\mat{\delta}_{p}\varphi\frac{\mathrm{d}p_{\mathrm{sat}}}{\mathrm{d}\theta}\vek\nabla\theta\right\}\right]\mathrm{d}\Omega +\nonumber \\
&& +\int_{\Omega}\left\{\vek{\nabla}\delta
\varphi\right\}^{\mathrm{\textsf{T}}}\left[h_{v}
\{\mat{\delta}_{p}
p_{\mathrm{sat}}\vek{\nabla}\varphi\}\right]\mathrm{d}\Omega -
\int_{\Gamma^{\bar{q}}_{\theta}}\delta
\theta\,\bar{q}_{\nu}\mathrm{d}\Gamma =0,
\label{eq:hom01}
\end{eqnarray}
\begin{eqnarray}
&&\int_{\Omega}\delta\varphi\frac{\mathrm{d}w}{\mathrm{d}\varphi}\frac{\mathrm{d}
\varphi}{\mathrm{d}t}\mathrm{d}\Omega +
+\int_{\Omega}\{\vek{\nabla}\delta
\varphi\}^{\mathrm{\textsf{T}}}\left[\{{D_{\varphi}}\vek{\nabla}\varphi\}+\{\mat{\delta}_{p}p_{\mathrm{sat}}\vek{\nabla}\varphi\}\right]\mathrm{d}\Omega +\nonumber \\
&& +\int_{\Omega}\left\{\vek{\nabla}\delta
\theta\right\}^{\mathrm{\textsf{T}}}\left\{\mat{\delta}_{p}\varphi\frac{\mathrm{d}p_{\mathrm{sat}}}{\mathrm{d}\theta}\vek\nabla\theta\right\}
\mathrm{d}\Omega
-\int_{\Gamma^{\bar{g}}_{\varphi}}\delta\varphi\,\bar{g}_{\nu}\mathrm{d}\Gamma = 0,
 \label{eq:DDE2}
\end{eqnarray}
to be solved numerically for the prescribed boundary and initial
conditions, see Fig.~\ref{fig:DDE1} for the definition of prescribed
boundary terms and domain representation. Owing to a strong nonlinear
dependence of material parameters on both temperature and moisture
fields, recall~\ref{app:A}, the Newton-Raphson method is generally needed
to solve the resulting discretized system of equations.

\section{First order homogenization of non-stationary coupled heat and moisture transport}\label{sec:Homogenization}
The present section derives the governing equations of the coupled
heat and moisture transport in the framework of coupled two-scale
analysis of FE$^2$ type. In this regard it is presumed that the
homogenized macro-scale fields are found from the solution of a
certain sub-scale (meso-scale) problem performed on a representative
volume element (RVE) being identical, at least in a statistical sense,
to a real meso-structure from both the geometrical and material
composition point of view. Such an RVE is then usually termed the
statistically equivalent periodic unit cell
(SEPUC)~\cite{Zeman:MSMSE:2007}. Examples of such SEPUCs for both
regular as well as irregular masonry walls adopted herein are plotted
in Fig.~\ref{fig:decomp}(d)(e). It has been advocated
in~\cite{Larsson:2010:IJMNE} that for a finite size RVE the assumption
of transient flow on both the macro and meso-scale introduces certain
non-local, size dependent, terms in equations governing the
macroscopic response. Some numerical simulations addressing this issue
are presented in Section~\ref{subsec:InfTRANS}, whilst the theoretical
grounds are provided next following closely, although in more
abbreviated format, the variationally consistent homogenization
outlined in detail in~\cite{Larsson:2010:IJMNE}.

To introduce this subject suppose that a local field $a$ can be
replaced by a spatially homogenized one $\langle{a}\rangle$ such that
\begin{eqnarray}
\int_{\Omega}a\,\mathrm{d}\Omega & \approx &
\int_{\Omega}\left\langle a\right\rangle_{\Box}
\mathrm{d}\Omega=\int_{\Omega}\left(\frac{1}{\left|\Omega_{\Box}\right|}\int_{\Omega_{\Box}}
a\,\mathrm{d}\Omega_{\Box}\right)\mathrm{d}\Omega, \label{eq:hom02} \\
\int_{\Gamma}a\,\mathrm{d}\Gamma & \approx &
\int_{\Gamma}\left\langle a\right\rangle_{\Box}
\mathrm{d}\Gamma=\int_{\Gamma}\left(\frac{1}{\left|\Gamma_{\Box}\right|}\int_{\Gamma_{\Box}}
a\,\mathrm{d}\Gamma_{\Box}\right)\mathrm{d}\Gamma,
\label{eq:hom03}
\end{eqnarray}
where $\Omega_{\Box}$ and $\Gamma_{\Box}$ represent the internal and
boundary parts of the SEPUC. In what follows, owing to the space
limitation, we shall treat only the energy balance
equation~\eqref{hom01} which upon employing Eqs. (\ref{eq:hom02}) and
(\ref{eq:hom03}) becomes
\begin{eqnarray}
&&\int_{\Omega}\left\langle\delta
\theta\frac{\mathrm{d}H}{\mathrm{d}\theta}\frac{\mathrm{d}\theta}{\mathrm{d}t}\right\rangle_{\Box}\mathrm{d}\Omega
+\int_{\Omega}\left\langle\{\vek{\nabla}\delta
\theta\}^{\mathrm{\textsf{T}}}\left[\{\mat{\lambda}\vek{\nabla}
\theta\}+h_{v}
\left\{\mat{\delta}_{p}\varphi\frac{\mathrm{d}p_{\mathrm{sat}}}{\mathrm{d}\theta}\vek\nabla\theta\right\}\right]\right\rangle_{\Box}\mathrm{d}\Omega +\nonumber \\
&& +\int_{\Omega}\left\langle\{\vek{\nabla}\delta
\varphi\}^{\mathrm{\textsf{T}}}\left[h_{v} \{\mat{\delta}_{p}
p_{\mathrm{sat}}\vek{\nabla}\varphi\}\right]\right\rangle_{\Box}\mathrm{d}\Omega
- \int_{\Gamma^{\bar{q}}_{\theta}}\left\langle\delta
\theta\,\bar{q}_{\nu}\right\rangle_{\Box}\mathrm{d}\Gamma =0.
\label{eq:hom04}
\end{eqnarray}

In the spirit of the first order homogenization it is assumed that the
macroscopic temperature and relative humidity vary only linearly over
the SEPUC. This can be achieved by loading its boundary by the
prescribed temperature $\Theta^{\mathrm{hom}}$ and relative humidity
$\Phi^{\mathrm{hom}}$ derived from the uniform macroscopic temperature
$\vek{\nabla}\Theta$ and relative humidity $\vek{\nabla}\Phi$
gradients. In such a case the local temperature and relative humidity
inside the SEPUC admit the following decomposition
\begin{eqnarray}
\theta(\vek{x}) & = & \Theta(\vek{X}^{0}) + \{\vek{\nabla}
\Theta\}^{\mathrm{\textsf{T}}}\{\vek{x} -
\vek{X}^{0}\}+\theta^{*}(\vek{x}) = \Theta^{\mathrm{hom}}(\vek{x})+\theta^{*}(\vek{x}), \label{eq:hom05} \\
\varphi(\vek{x}) & = & \Phi(\vek{X}^{0}) + \{\vek{\nabla}
\Phi\}^{\mathrm{\textsf{T}}}\{\vek{x} -
\vek{X}^{0}\}+\varphi^{*}(\vek{x})\, =
\Phi^{\mathrm{hom}}(\vek{x}) + \varphi^{*}(\vek{x}),
\label{eq:hom06}
\end{eqnarray}
where $\theta^{*}(\vek{x})$ and $\varphi^{*}(\vek{x})$ are the
fluctuations of local fields superimposed onto linearly varying
quantities $\Theta^{\mathrm{hom}}(\vek{x})$ and
$\Phi^{\mathrm{hom}}(\vek{x})$ . The temperature $\Theta(\vek{X}^{0})$
and the moisture $\Phi(\vek{X}^{0})$ at the reference point
$\vek{X}^{0}$ are introduced to link the local fields to their
macroscopic counterparts. For convenience the SEPUC is typically
centered at $\vek{X}^{0}$. Henceforth, the local fluctuations will be
demanded to be periodic, i.e. the same values are enforced on the
opposite sides of a rectangular SEPUC.

Next, substituting Eqs. (\ref{eq:hom05}) and (\ref{eq:hom06}) into
Eq. (\ref{eq:hom04}) and collecting the terms corresponding to
$\delta\Theta^{\mathrm{hom}}$, $\delta\Phi^{\mathrm{hom}}$ and
$\delta\theta^*$, $\delta\varphi^*$ splits the original
problem~\eqref{hom01} into the homogenized (macro-scale) problem
\begin{eqnarray}
&&\int_{\Omega}\left\langle \{\delta
\Theta^{\mathrm{hom}}\}\frac{\mathrm{d}H}{\mathrm{d}\theta}\frac{\mathrm{d}\theta}{\mathrm{d}t}\right\rangle_{\Box}\mathrm{d}\Omega
+\int_{\Omega}\left\langle\{\vek{\nabla}\delta
\Theta^{\mathrm{hom}}\}^{\mathrm{\textsf{T}}}\left[\{\mat{\lambda}\vek{\nabla}
\theta\}+h_{v}
\left\{\mat{\delta}_{p}\varphi\frac{\mathrm{d}p_{\mathrm{sat}}}{\mathrm{d}\theta}\vek\nabla\theta\right\}\right]\right\rangle_{\Box}\mathrm{d}\Omega +\nonumber \\
&& +\int_{\Omega}\left\langle\{\vek{\nabla}\delta
(\Phi^{\mathrm{hom}})\}^{\mathrm{\textsf{T}}}\left[h_{v}
\{\mat{\delta}_{p}
p_{\mathrm{sat}}\vek{\nabla}\varphi\}\right]\right\rangle_{\Box}\mathrm{d}\Omega
- \int_{\Gamma^{\bar{q}}_{\theta}}\left\langle \{\delta
\Theta^{\mathrm{hom}}\}\,\bar{q}_{\nu}\right\rangle_{\Box}\mathrm{d}\Gamma = 0,
\label{eq:hom07-1}
\end{eqnarray}
and the local sub-scale (meso-scale) problem
\begin{eqnarray}
&&\int_{\Omega}\left\langle
\{\delta\theta^{*}\}\frac{\mathrm{d}H}{\mathrm{d}\theta}\frac{\mathrm{d}\theta}{\mathrm{d}t}\right\rangle_{\Box}\mathrm{d}\Omega
+\int_{\Omega}\left\langle\{\vek{\nabla}\delta
\theta^{*}\}^{\mathrm{\textsf{T}}}\left[\{\mat{\lambda}\vek{\nabla}
\theta\}+h_{v}
\left\{\mat{\delta}_{p}\varphi\frac{\mathrm{d}p_{\mathrm{sat}}}{\mathrm{d}\theta}\vek\nabla\theta\right\}\right]\right\rangle_{\Box}\mathrm{d}\Omega +\nonumber \\
&& +\int_{\Omega}\left\langle\{\vek{\nabla}\delta
\varphi^{*}\}^{\mathrm{\textsf{T}}}\left[h_{v} \{\mat{\delta}_{p}
p_{\mathrm{sat}}\vek{\nabla}\varphi\}\right]\right\rangle_{\Box}\mathrm{d}\Omega
- \underbrace{\int_{\Gamma^{\bar{q}}_{\theta}}\left\langle
\{\delta\theta^{*}\}\,\bar{q}_{\nu}\right\rangle_{\Box}\mathrm{d}\Gamma}_{=0\;\;{\rm due\; to\; periodicity}}
= \label{eq:hom07-2}\\
&&\int_{\Omega}\frac{1}{\left|\Omega_{\Box}\right|}\int_{\partial\Omega^{\bar{q}}_{\Box}}
\{\delta
\theta^{*}\}\,\bar{q}_{\nu}\,\mathrm{d}(\partial\Omega_{\Box})\,\mathrm{d}\Omega=0,\nonumber
\end{eqnarray}
which is satisfied identically owing to the assumed periodic boundary
conditions. Solving Eq.~\eqref{hom07-2} for the prescribed increments
of $\vek{\nabla}\Theta$ and $\vek{\nabla}\Phi$ provides the
instantaneous effective properties and storage terms that appear in
the macro-scale equation~\eqref{hom07-1}. Because of a strong
non-linearity the two equations must be solved iteratively in a
certain nested loop, see~\cite{Ozdemir:IJNME:2008,Larsson:2010:IJMNE}
for further reference.

Since details on the solution of Eq.~\eqref{hom07-2} are available in
our preceding paper~\cite{Sykora:2011:AMC} we limit our attention to
the macro-scale problem and write the first term of
Eq.~\eqref{hom07-1} with the help of Eqs. (\ref{eq:hom05}) and
(\ref{eq:hom06}) as
\begin{eqnarray}
&&\int_{\Omega}\left\langle \{\delta
\Theta^{\mathrm{hom}}\}\frac{\mathrm{d}H}{\mathrm{d}\theta}\frac{\mathrm{d}\theta}{\mathrm{d}t}\right\rangle_{\Box}\mathrm{d}\Omega=
\int_{\Omega}\left\langle \{\delta(\Theta + \{\vek{\nabla}
\Theta\}^{\mathrm{\textsf{T}}}\{\vek{x} -
\vek{X}^{0}\})\}\frac{\mathrm{d}H}{\mathrm{d}\theta}\frac{\mathrm{d}\theta}{\mathrm{d}t}\right\rangle_{\Box}\mathrm{d}\Omega=\nonumber\\
&&=\int_{\Omega}\left\langle
\{\delta\Theta\}\frac{\mathrm{d}H}{\mathrm{d}\theta}\frac{\mathrm{d}\theta}{\mathrm{d}t}+
\{\delta\vek{\nabla} \Theta\}^{\mathrm{\textsf{T}}}\{\vek{x}-
\vek{X}^{0}\}\frac{\mathrm{d}H}{\mathrm{d}\theta}\frac{\mathrm{d}\theta}{\mathrm{d}t}\right\rangle_{\Box}\mathrm{d}\Omega,
\label{eq:hom09}
\end{eqnarray}
thus clearly identifying the solution dependence on the actual size of
the SEPUC through the second term in the integral~\eqref{hom09}. We may
now substitute from Eq.~\eqref{hom09} into Eq.~\eqref{hom07-1} to get
\begin{eqnarray}
&&-\underbrace{\int_{\Omega}\left\langle\{\delta
\Theta\}^{\mathrm{\textsf{T}}}
\frac{\mathrm{d}H}{\mathrm{d}\theta}\frac{\mathrm{d}\theta}{\mathrm{d}t}\right\rangle_{\Box}\mathrm{d}\Omega}_{\mathrm{C}_{\theta\theta}\frac{\mathrm{d}r_{\theta}}{\mathrm{d}t}}
-\underbrace{\int_{\Omega}\left\langle \{\delta\vek{\nabla}
\Theta\}^{\mathrm{\textsf{T}}} \{\vek{x}
-\vek{X}^{0}\}\frac{\mathrm{d}H}{\mathrm{d}\theta}\frac{\mathrm{d}\theta}{\mathrm{d}t}\right\rangle_{\Box}\mathrm{d}\Omega}_{\mathrm{C}_{\theta\theta}^{'}\frac{\mathrm{d}r_{\theta}}{\mathrm{d}t}}
- \nonumber \\
&& -\underbrace{\int_{\Omega}\left\langle\{\delta\vek{\nabla}
\Theta \}^{\mathrm{\textsf{T}}}\left[\{\mat{\lambda}\vek{\nabla}
\theta\}+h_{v}
\left\{\mat{\delta}_{p}\varphi\frac{\mathrm{d}p_{\mathrm{sat}}}{\mathrm{d}\theta}\vek\nabla\theta\right\}\right]\right\rangle_{\Box}\mathrm{d}\Omega}_{\mathrm{K}_{\theta\theta}r_{\theta}} -\nonumber \\
&& -\underbrace{\int_{\Omega}\left\langle\{\delta \vek{\nabla}
\Phi \}^{\mathrm{\textsf{T}}}\left[h_{v} \{\mat{\delta}_{p}
p_{\mathrm{sat}}\vek{\nabla}\varphi\}\right]\right\rangle_{\Box}\mathrm{d}\Omega}_{\mathrm{K}_{\theta\varphi}r_{\varphi}}
+ \underbrace{\int_{\Gamma^{\bar{q}}_{\theta}}\left\langle\{\delta
\Theta\}^{\mathrm{\textsf{T}}}
\,\bar{q}_{\nu}\right\rangle_{\Box}\mathrm{d}\Gamma}_{q_{\mathrm{ext}}}
=0.\label{eq:hom10}
\end{eqnarray}
An analogous approach can be applied also to the moisture transport
equation~\eqref{DDE2} to arrive, after classical finite element
discretization, into a discretized system of coupled macroscopic heat
and moisture equations
\begin{eqnarray}
\mat{K}_{\theta\theta}\vek{r}_{\theta}+\mat{K}_{\theta\varphi}\vek{r}_{\varphi}+(\mat{C}_{\theta\theta}+\mat{C}_{\theta\theta}^{'})\frac{\mathrm{d}\vek{r}_{\theta}}{\mathrm{d}t}
& = & \vek{q}_{\mathrm{ext}}, \label{eq:hom11}\\
\mat{K}_{\varphi\theta}\vek{r}_{\theta}+\mat{K}_{\varphi\varphi}\vek{r}_{\varphi}+(\mat{C}_{\varphi\varphi}+\mat{C}_{\varphi\varphi}^{'})\frac{\mathrm{d}\vek{r}_{\varphi}}{\mathrm{d}t}
& = & \vek{g}_{\mathrm{ext}}, \label{eq:hom12}
\end{eqnarray}
which have to be properly integrated in the time domain adopting for
example the Crank-Nicolson integration scheme. Details on the
numerical implementation are available in~\cite{Sykora:2010}.

\section{Examples}\label{sec:Examples}
Several illustrative example problems were analyzed to address the
non-linear transient coupled heat and moisture transport assumed on
both scales, the influence of the way of prescribing the macroscopic
loading conditions closely related to the macro-scale finite element
mesh and finally the solution strategy exploiting the parallel
computation. The same material data were adopted in all analyses.
These were obtained from a set of experimental measurements providing
the hygric and thermal properties of mortars and bricks/stones, which
have been used in the reconstructions works of historical buildings in
the Czech Republic including Charles Bridge,
see~\cite{Pavlikova:NSBP:2008}. The measured material parameters of
individual masonry phases listed in Table.~\ref{tab:matpar} then
served to derive the non-measurable transport coefficients presented
in Eqs.~(\ref{eq:Kun01}) and (\ref{eq:Kun02}), see also~\ref{app:A}.

\begin{table}[h!]
\begin{center}
\begin{tabular}{lllcc}
\multicolumn{3}{l}{parameter} & brick & mortar \\
\hline
$w_{f}$ &  $\mathrm{[kgm^{-3}]}$ &free water saturation & 229.30 & 160.00 \\
$w_{\mathrm{80}}$  & $\mathrm{[kgm^{-3}]}$ & water content at $\varphi=0.8$ [-] & 141.68 & 22.72 \\
$\lambda_{\mathrm{0}}$ & $\mathrm{[Wm^{-1}K^{-1}]}$ & thermal conductivity & 0.25 & 0.45 \\
$b_{\mathrm{tcs}}$  & $\mathrm{[-]}$ & thermal conductivity supplement & 10 & 9  \\
$\rho_{s}$ & $\mathrm{[kgm^{-3}]}$ &  bulk density & 1690 & 1670 \\
$\mu$   & $\mathrm{[-]}$ &water vapor diffusion resistance & 16.80 & 9.63 \\
$A$ & $\mathrm{[kgm^{-2}s^{-0.5}]}$ &  water absorption coefficient & 0.51 & 0.82 \\
$c_s$ & $\mathrm{[Jkg^{-1}K^{-1}]}$ &specific heat capacity & 840 & 1000 \\
\hline
\end{tabular}
\caption{Material parameters of individual phases.}
\label{tab:matpar}
\end{center}
\end{table}

\subsection{Influence of transient flow at meso-level}\label{subsec:InfTRANS}
This section supports through numerical simulations the theoretically
predicted size dependence of the homogenized response first suggested
for the case of a non-linear single variable diffusion problem
in~\cite{Larsson:2010:IJMNE} and also established here in
Section~\ref{sec:Homogenization} for the special case of the coupled
heat and moisture problem in the framework of K\"{u}nzel's
constitutive model.

\begin{figure} [h!]
\begin{center}
\begin{tabular}{c@{\hspace{5mm}}c}
\includegraphics*[width=65mm,keepaspectratio]{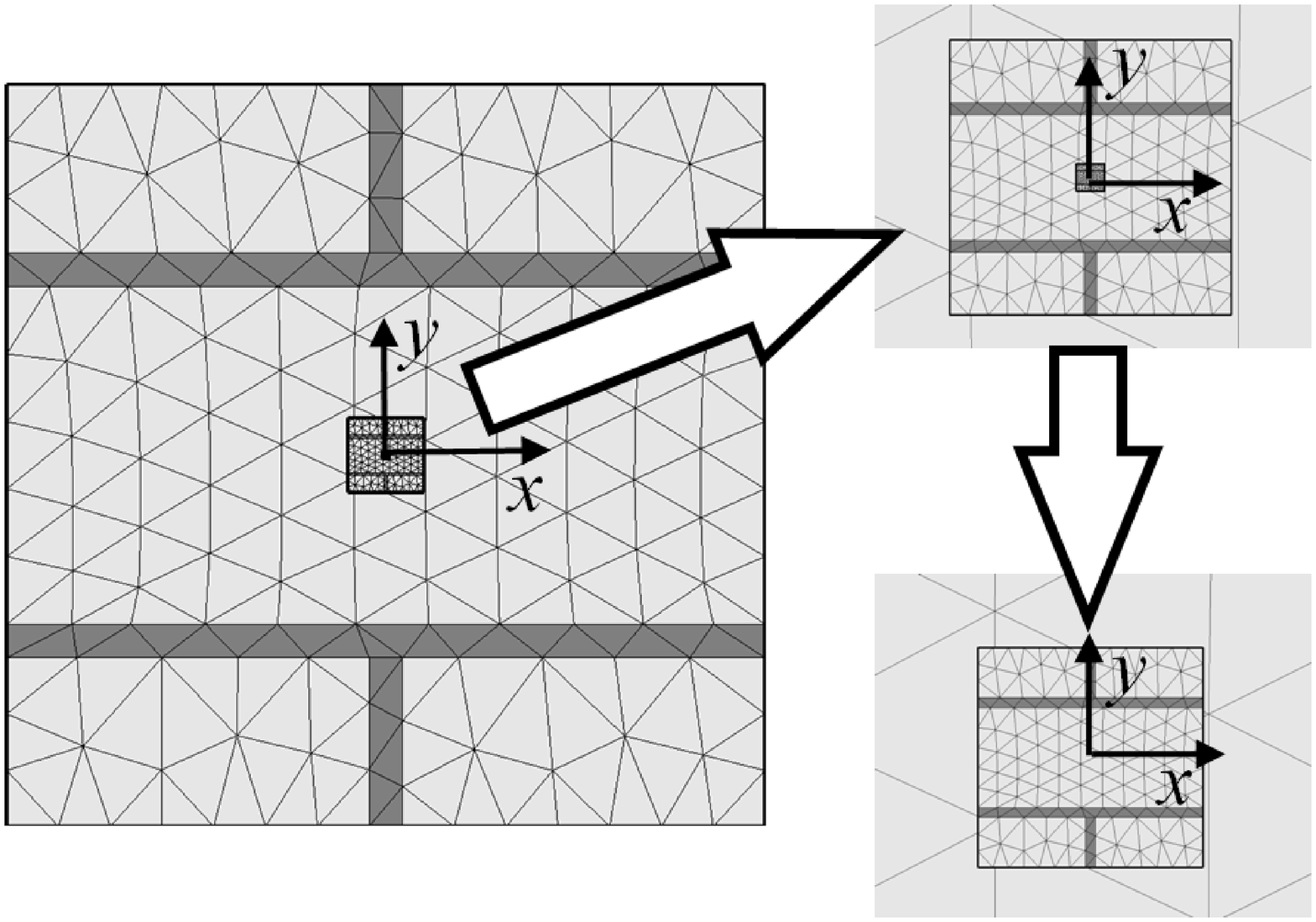}&
\includegraphics*[width=65mm,keepaspectratio]{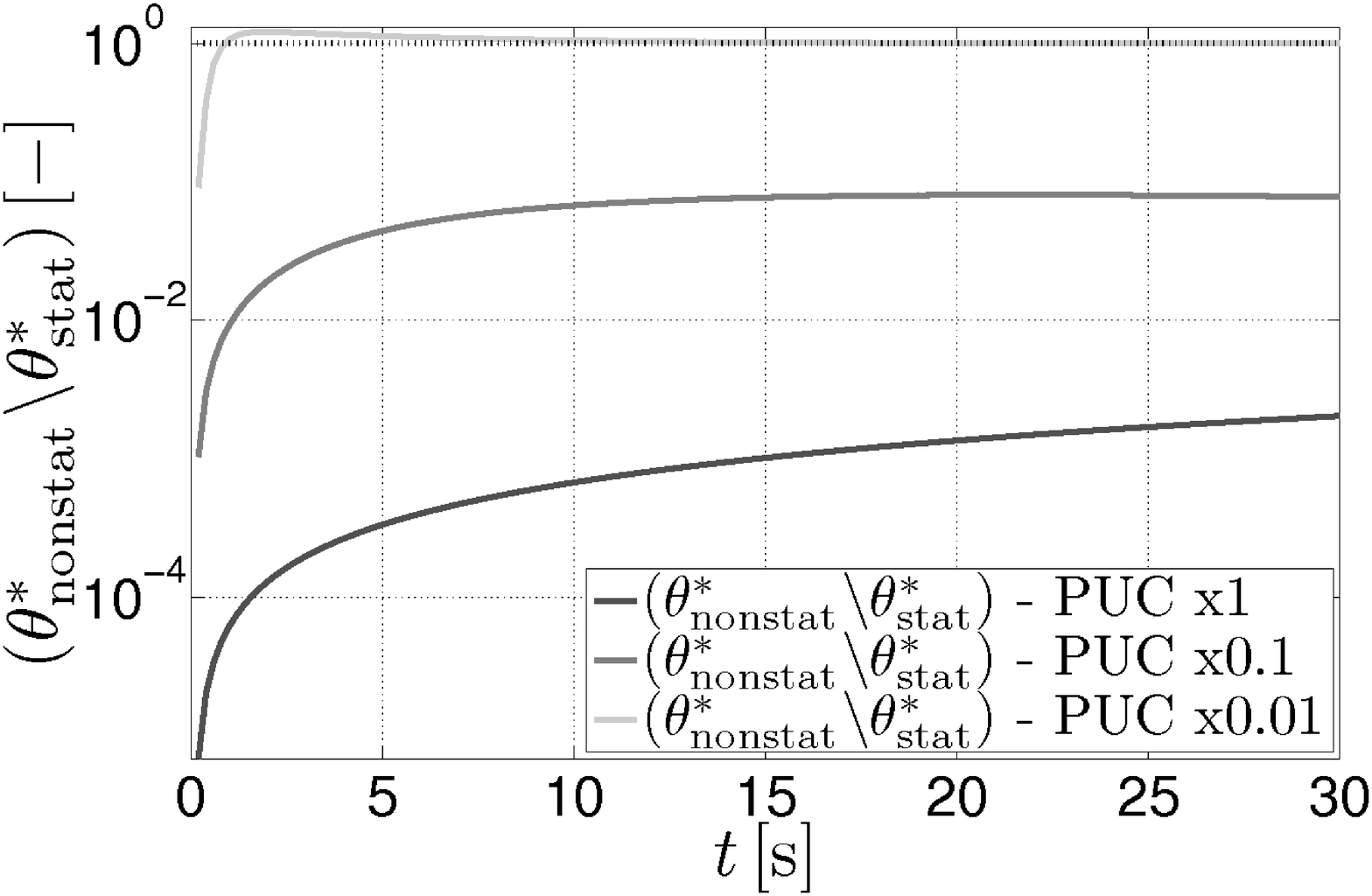}\\
(a)&(b)
\end{tabular}
\end{center}
\caption{(a) Investigated periodic unit cells, (b) resulting evolution of fluctuation temperature}
\label{fig:PUC}
\end{figure}

In doing so we considered three particular units cells in
Fig.~\ref{fig:PUC}(a) varying in size from millimeters to decimeters.
Each cell was loaded by the same constant gradients of temperature
and moisture along the $x$-direction. The resulting evolutions of the
fluctuation part of the local temperature at the center of individual
cells appear in Fig.~\ref{fig:PUC}(b) clearly manifesting the
influence of the size of the cell which necessary projects into the
prediction of the homogenized properties and thus evolution of the
predicted macroscopic response. For the largest cell the steady state
was reached in about $80$ [h].

\subsection{Influence of macrostructural finite element mesh}\label{subsec:InfFEM}
One of the concerns of the implementation of FE$^2$ scheme is the
application of boundary conditions on the macro-scale keeping in mind
the scale transition requirement and periodic boundary conditions
imposed on the meso-scale. This may become important particularly with
a relatively large periodic unit cells which may even exceed the size
of macroscopic elements in the vicinity of outer boundary where the
macro-elements should be fine enough to ensure a smooth and accurate
evolution of driving variables from the outside into the inner parts
of a structure.

\begin{figure} [h!]
\begin{center}
\begin{tabular}{ccc}
\includegraphics*[width=42mm,keepaspectratio]{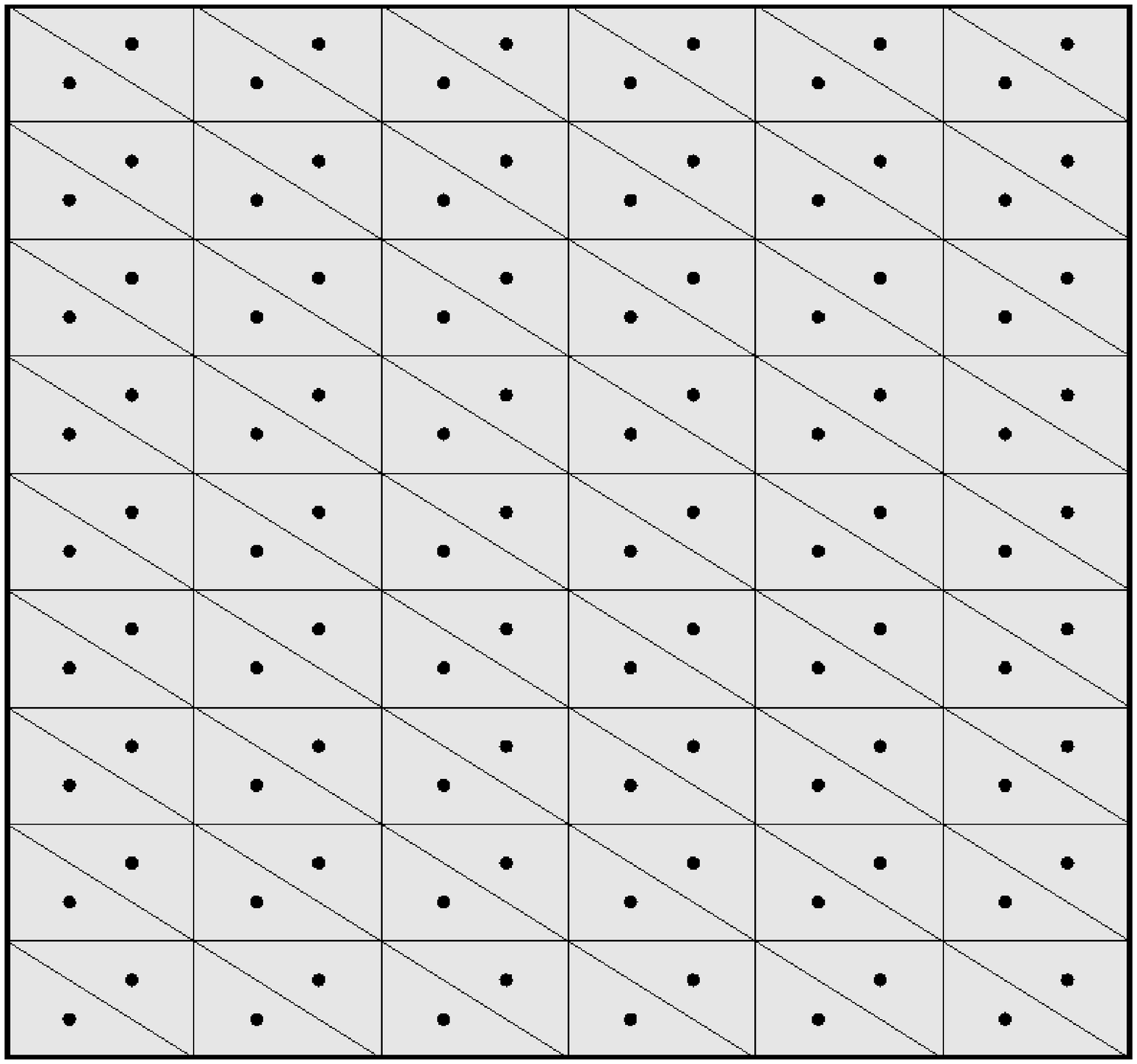}&
\includegraphics*[width=42mm,keepaspectratio]{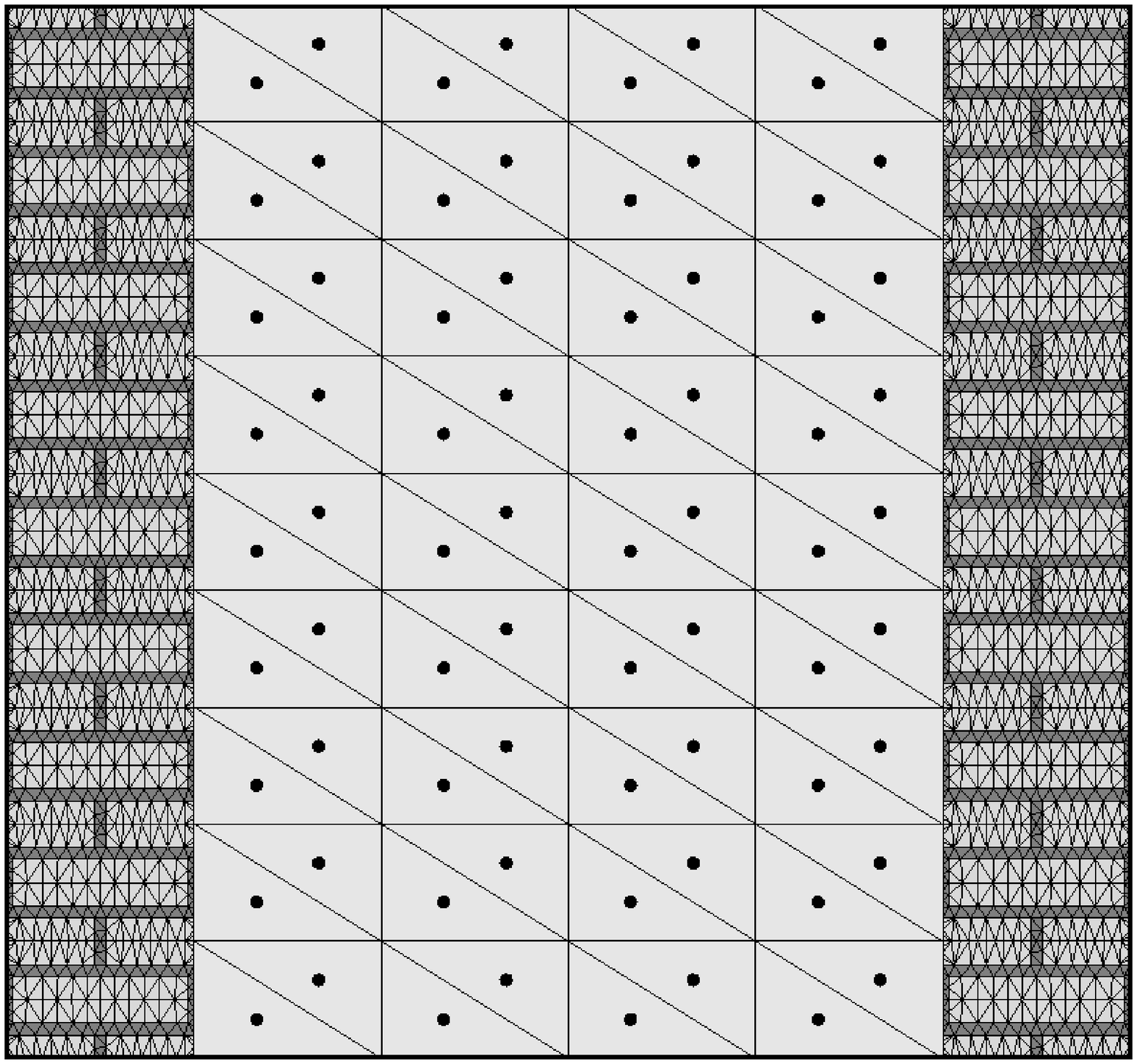}&
\includegraphics*[width=42mm,keepaspectratio]{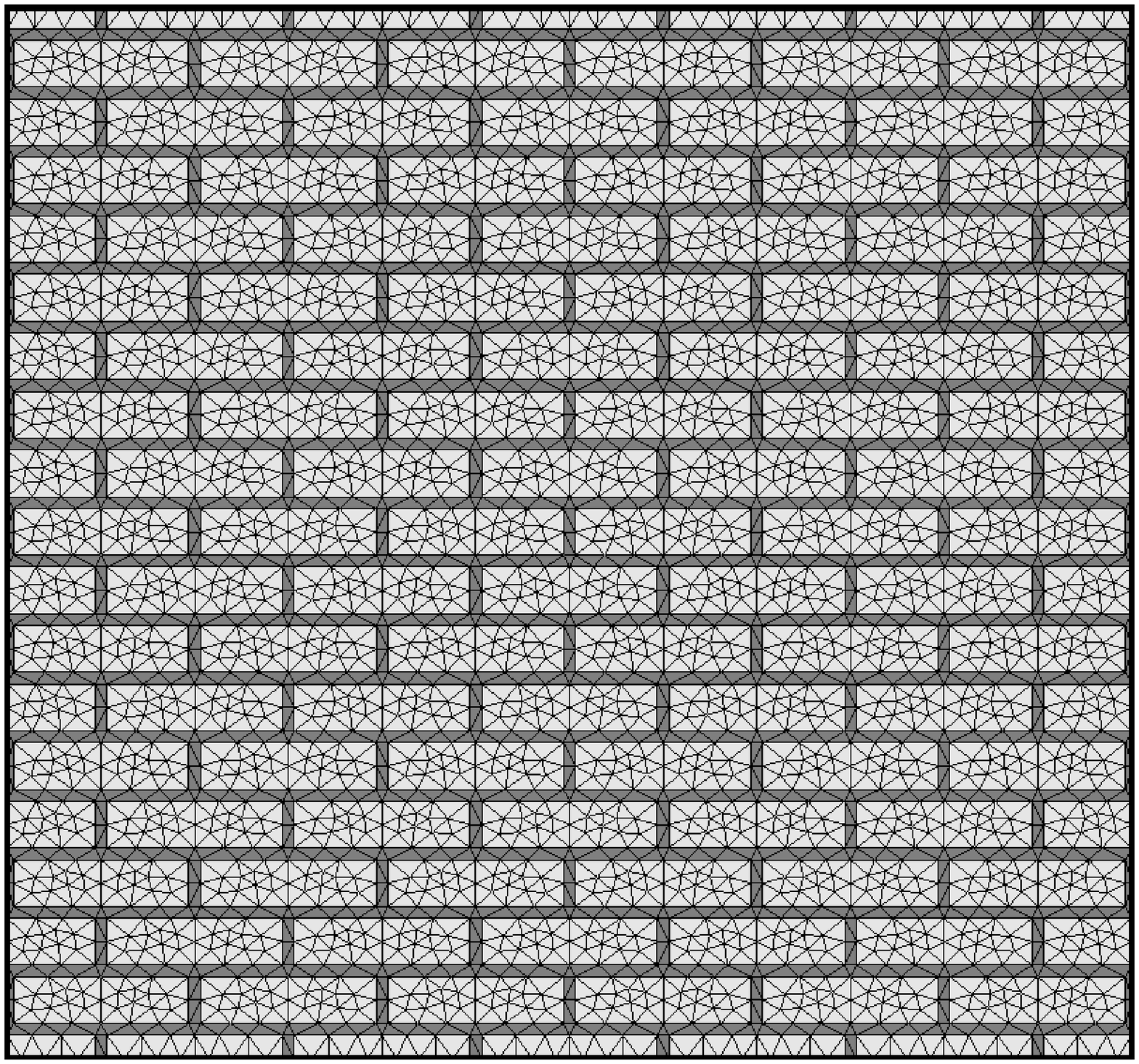}\\
(a)&(b)&(c)
\end{tabular}
\end{center}
\caption{Different finite element representations of masonry wall
($l_{x}=1.92\,\mathrm{[m]}$, $l_{y}=1.80\,\mathrm{[m]}$) -
(a) full multi-scale scheme,
(b) semi multi-scale scheme,
(c) full fine-scale discretization}
\label{fig:mesh}
\end{figure}

To address this issue we studied two types of macro-scale
discretizations adopted in the two-scale (meso-macro) analysis.  The
corresponding macro-scale finite element meshes appear in
Figs.~\ref{fig:mesh}(a),(b). The case when the fine-scale details are
assigned to the entire structure is plotted in Fig.~\ref{fig:mesh}(c).
This mesh, consisting of $7050$ triangular finite elements, served to
evaluate the accuracy of the two former
discretizations. Fig.~\ref{fig:mesh}(a) shows $108$ macro-elements
each representing a single meso-problem with assigned periodic
boundary conditions, whereas the case in Fig.~\ref{fig:mesh}(b)
assumes the outer boundary being fully discretized (note that only
one-directional flow is considered). There, only the inner part
consisting of 72 elements is subject to multi-scale analysis whilst
the outer part is modeled as a structure with a real masonry bonding
consisting of $3888$ finite elements. A multi-point constrains were
introduced to account for an incompatible discretization along the
common interface. This latter case is, therefore, expected to heal
inaccuracies in the estimation of temperature and moisture fields in
the region close to the surface layer.

\begin{figure} [h!]
\begin{center}
\begin{tabular}{c@{\hspace{5mm}}c}
\includegraphics*[width=65mm,keepaspectratio]{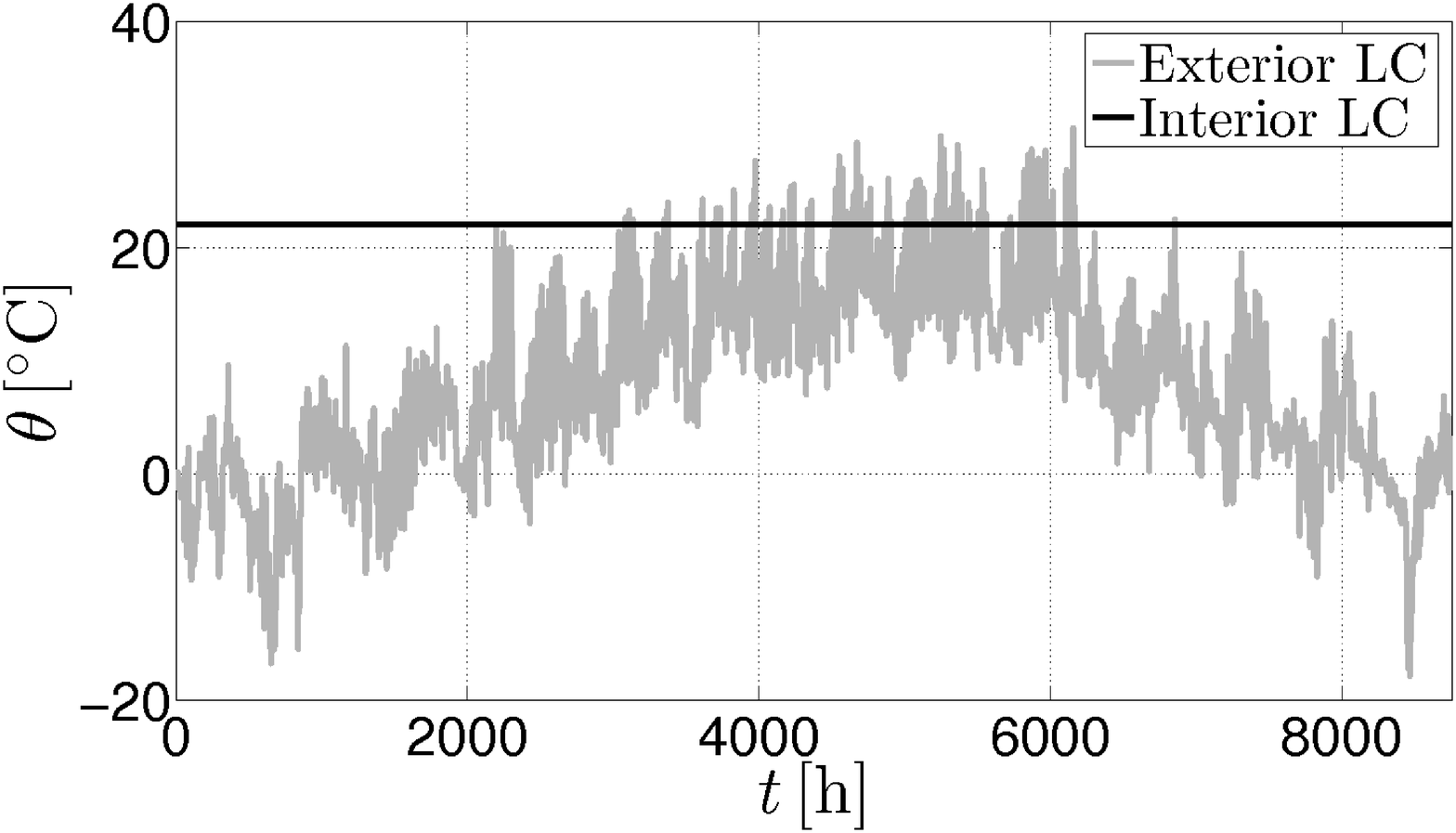}&
\includegraphics*[width=65mm,keepaspectratio]{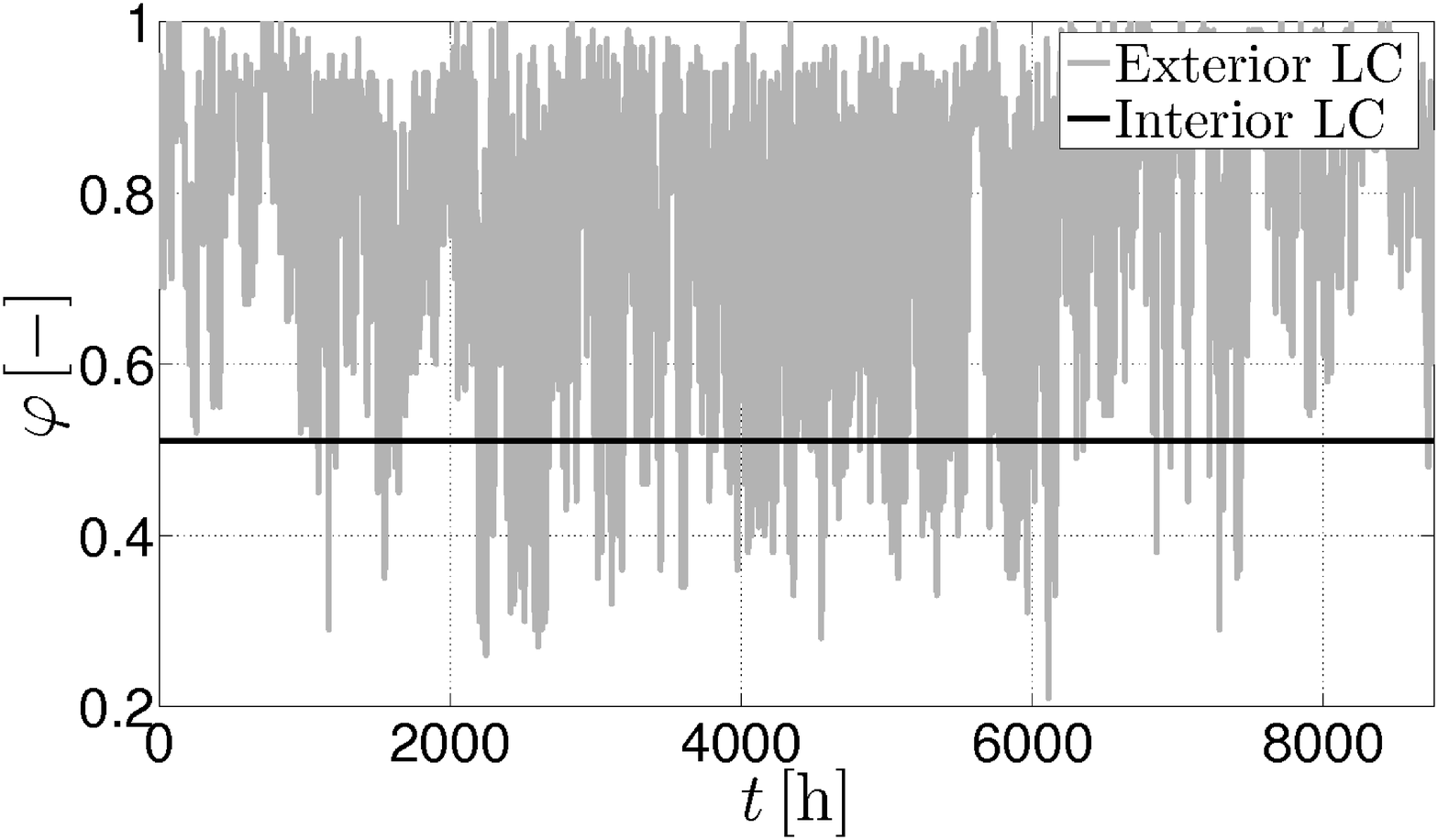}\\
(a)&(b)
\end{tabular}
\end{center}
\caption{Annual loading conditions - (a) temperature, (b) moisture}
\label{fig:BC}
\end{figure}

The following boundary conditions were imposed: on the right-hand side
(interior) a constant temperature of $24\,\mathrm{[^{\circ}C]}$ and a
constant relative humidity $0.5\,\mathrm{[-]}$ were maintained, while
on the left-hand side (exterior) the real climatic data collected over
the entire year were prescribed, see Fig.~\ref{fig:BC}.

\begin{figure} [h!]
\begin{center}
\begin{tabular}{c@{\hspace{5mm}}c}
\includegraphics*[width=65mm,keepaspectratio]{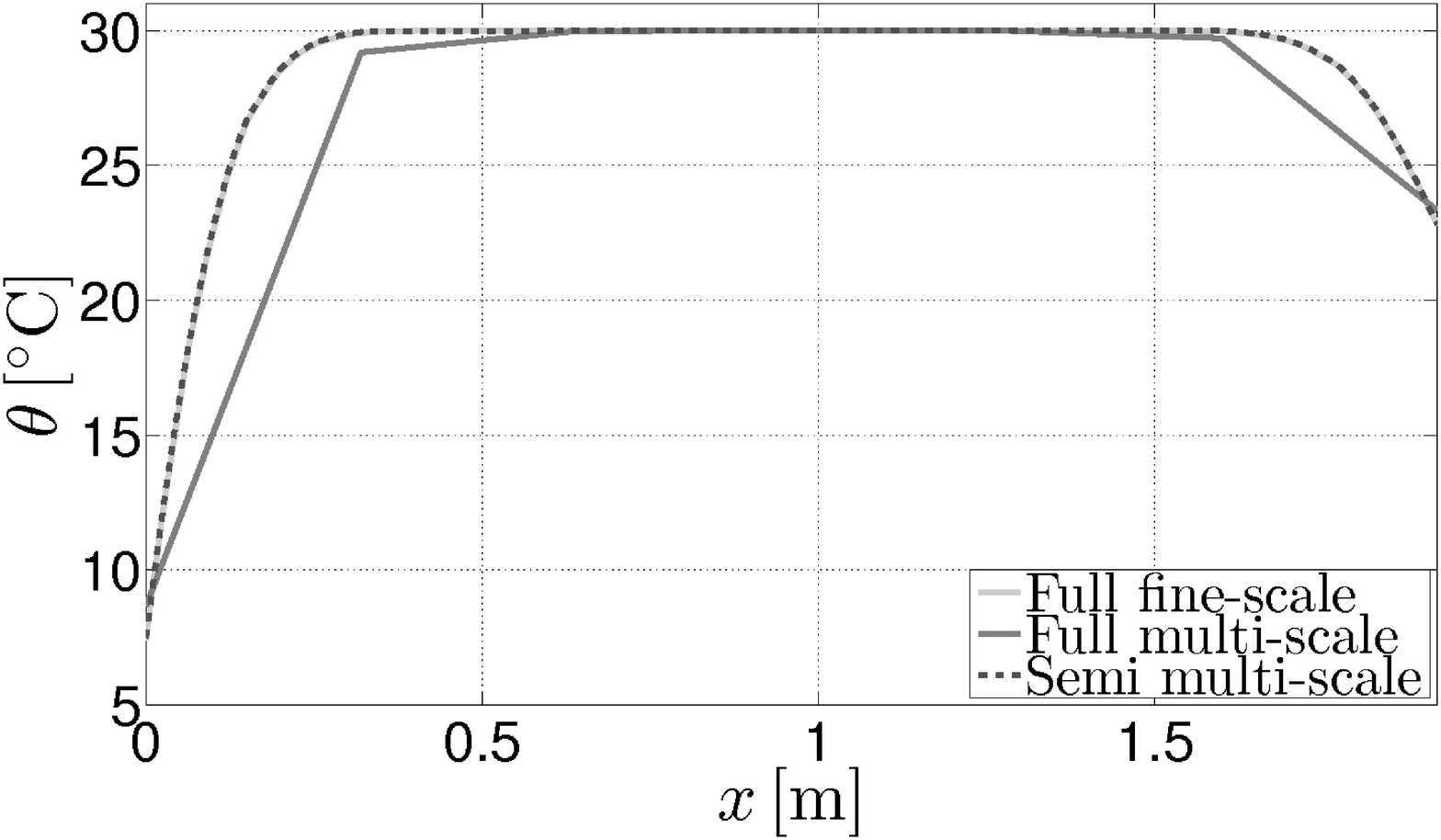}&
\includegraphics*[width=65mm,keepaspectratio]{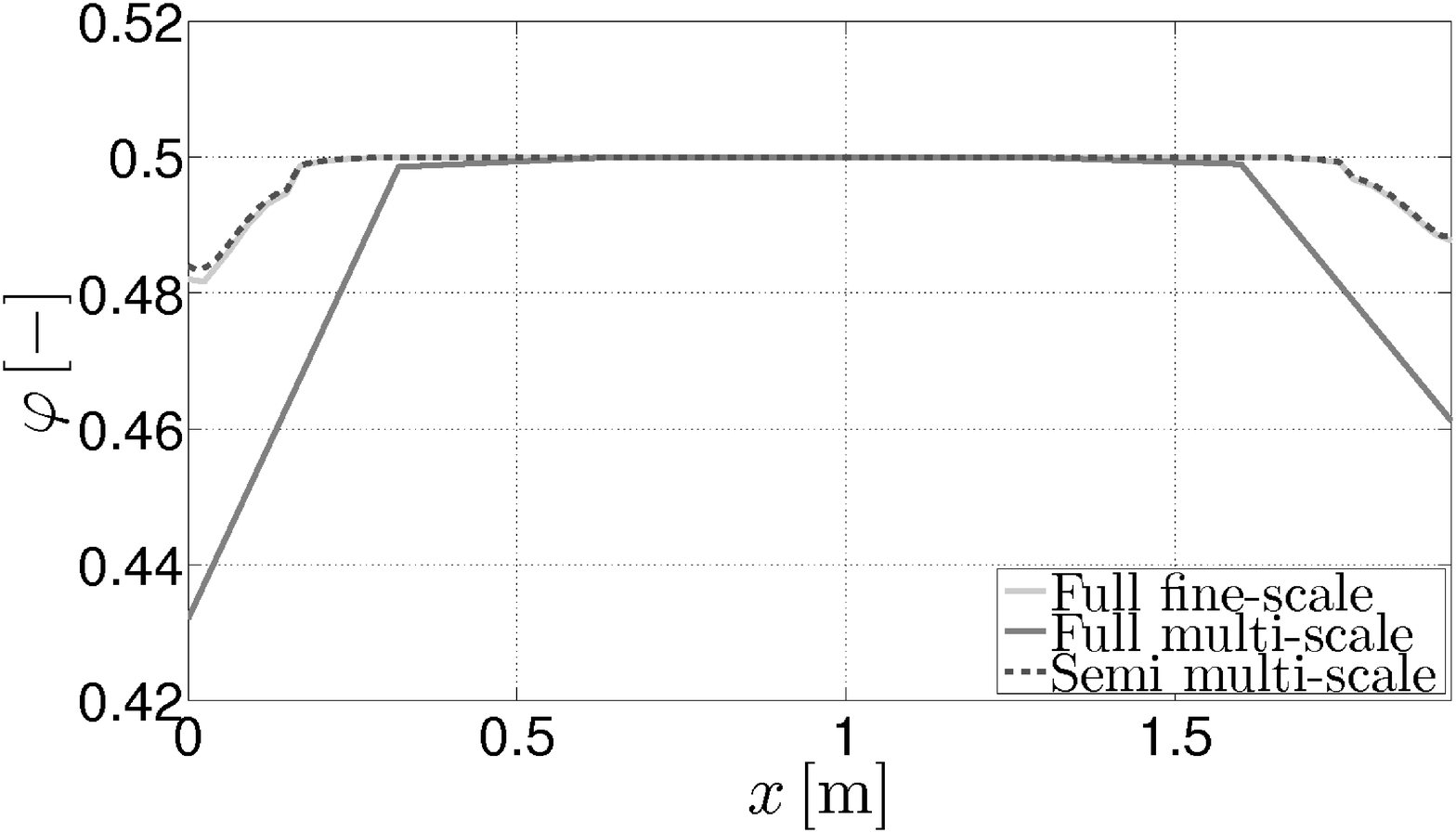}\\
(a)&(b)\\
\includegraphics*[width=65mm,keepaspectratio]{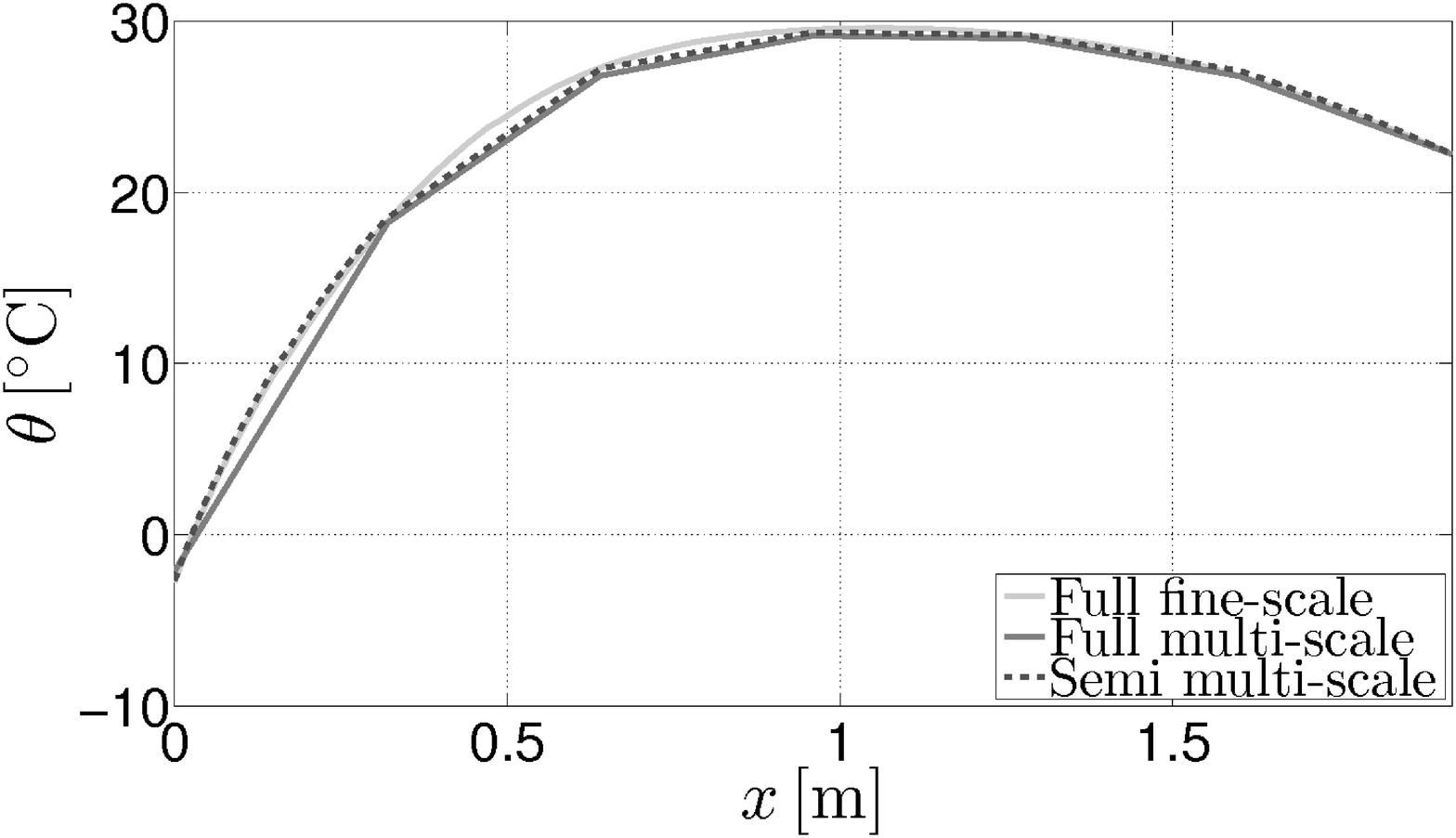}&
\includegraphics*[width=65mm,keepaspectratio]{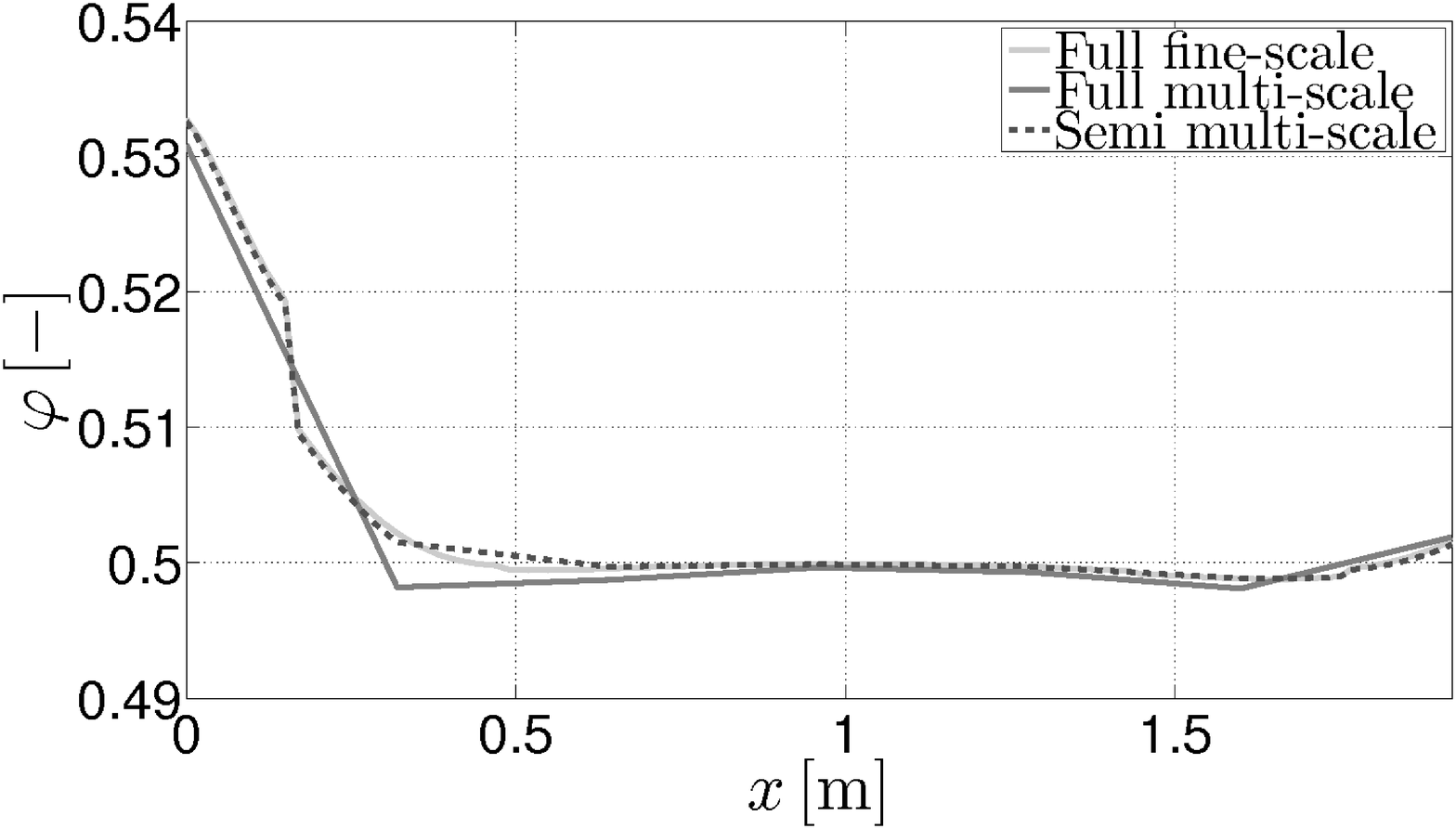}\\
(c)&(d)
\end{tabular}
\end{center}
\caption{Comparison of different macrostructural computations -
(a) temperature profile after $t=10\,\mathrm{[h]}$,
(b) moisture profile after $t=10\,\mathrm{[h]}$
(c) temperature profile after $t=100\,\mathrm{[h]}$,
(d) moisture profile after $t=100\,\mathrm{[h]}$}
\label{fig:Comp}
\end{figure}

The results appear in Fig.~\ref{fig:Comp} showing variation of the
temperature and moisture along the mid section of the wall after the
duration of load of $10$ [h] and $100$ [h], respectively, derived for
the macroscopic time step equal to $1$ [h]. Clearly, the notable
difference between the exact (full fine-scale discretization) and full
multi-scale scheme can be observed in surface layers only and this
difference almost disappears with a sufficiently long duration of
time. It thus appears that the refined representation of the surface
layer through the semi multi-scale scheme, although more accurate
compare to full multi-scale scheme, does not bring any particular
advantage. This is supported by the calculated average and absolute
errors stored in Table~\ref{tab:Comp} taking into account all nodal
macroscopic temperatures and moistures in the domain over all time
integration steps. Note that for the sake of comparison the fine-scale
variables (solution employing the mesh in Fig.~\ref{fig:mesh}(c)) were
averaged over the cell basically covered by two macro-elements in
Fig.~\ref{fig:mesh}(a).

The presented results thus promote the more accurate semi multi-scale
scheme only for calculations demanding higher accuracy of local
results especially in initial stages of computation and/or examples
with fast changing boundary conditions.

\begin{table}[h!]
\begin{center}
\begin{tabular}{lcc}
type of comparison & avg. relative error & avg. absolute error \\
& $[\%]$ & $\mathrm{[^{\circ}C]}/[-]$ \\
\hline
calculation of temperature & & \\
- fine-scale vs. multi-scale & $2.62$ & $0.11$\\
- fine-scale vs. semi multi-scale & $0.71$ & $0.03$\\
calculation of relative humidity & & \\
- fine-scale vs. multi-scale & $0.18$ & $0.001$\\
- fine-scale vs. semi multi-scale & $0.05$ & $0.001$\\
\hline
\end{tabular}
\end{center}
\caption{Averaged relative and absolute errors}
\label{tab:Comp}
\end{table}

\subsection{Parallel computation}\label{subsec:Parallelization}
The essential request by the contractor when studying the mechanical
response of Charles Bridge to provide the basis for reconstruction
works was a full scale three-dimensional analysis of the
bridge. Performing such an analysis in a fully coupled format on a
single computer would be computationally unfeasible thus creating the
need for a parallel computing. Concentrating on the implementation
part of the parallel version of FE$^2$ scheme we limit our attention
to a two-dimensional section of Charles Bridge subjected, however, to
real climatic data displayed already in Fig.~\ref{fig:BC}. Extension
to a fully three-dimensional problem is under current investigation
and will be presented elsewhere.

As already discussed in the previous section, the present FE$^2$ based
multi-scale analysis assumes each macroscopic integration point be
connected with a certain mesoscopic problem represented by an
appropriate periodic unit cell. The solution of a meso-scale problem
then provides instantaneous effective data needed on the
macro-scale. Such an analysis is particularly suitable for a parallel
computing because the amount of transferred data is small. In this
regard, the master-slave strategy can be efficiently exploited. To
that end, the macro-problem is assigned to the master processor while
the solution at the meso-level is carried out on slave processors. At
each time step the current temperature and moisture together with the
increments of their gradients at a given macroscopic integration point
are passed to the slave processor (imposed onto the associated
periodic cell), which, upon completing the small scale analysis, sends
the homogenized data (effective conductivities, averaged storage terms
and fluxes) back to the master processor.

If the meso-scale problems are large enough, the ideal solution is to
assign one meso-problem to one slave processor. Clearly, even for very
small macro-problems with a few thousands of finite elements, the
hardware requirements would be in such a case excessive. On the other
hand, if the meso-problems are relatively small, i.e. they contain
small number of finite elements, the corresponding analysis might be
even shorter than the data transfer between the processors. Then, the
computational time associated with the data transfer between the
master processor and many slave processors may grow excessively.  It
is worth mentioning that this time consists of two contributions. The
first one represents the latency time (the processors make connection)
which is independent of the amount of transferred data whilst the
second contribution clearly depends on the amount of data being
transferred. For small meso-problems it is therefore reasonable to
assign several of them to a single slave processor. The master
processor then sends a larger package of data from many macroscopic
integration points at the same time to each slave processor so that
the latency time does not play a crucial role. This approach was
adopted hereinafter.

\begin{figure} [h!]
\begin{center}
\begin{tabular}{cc}
\includegraphics[width=65mm,keepaspectratio]{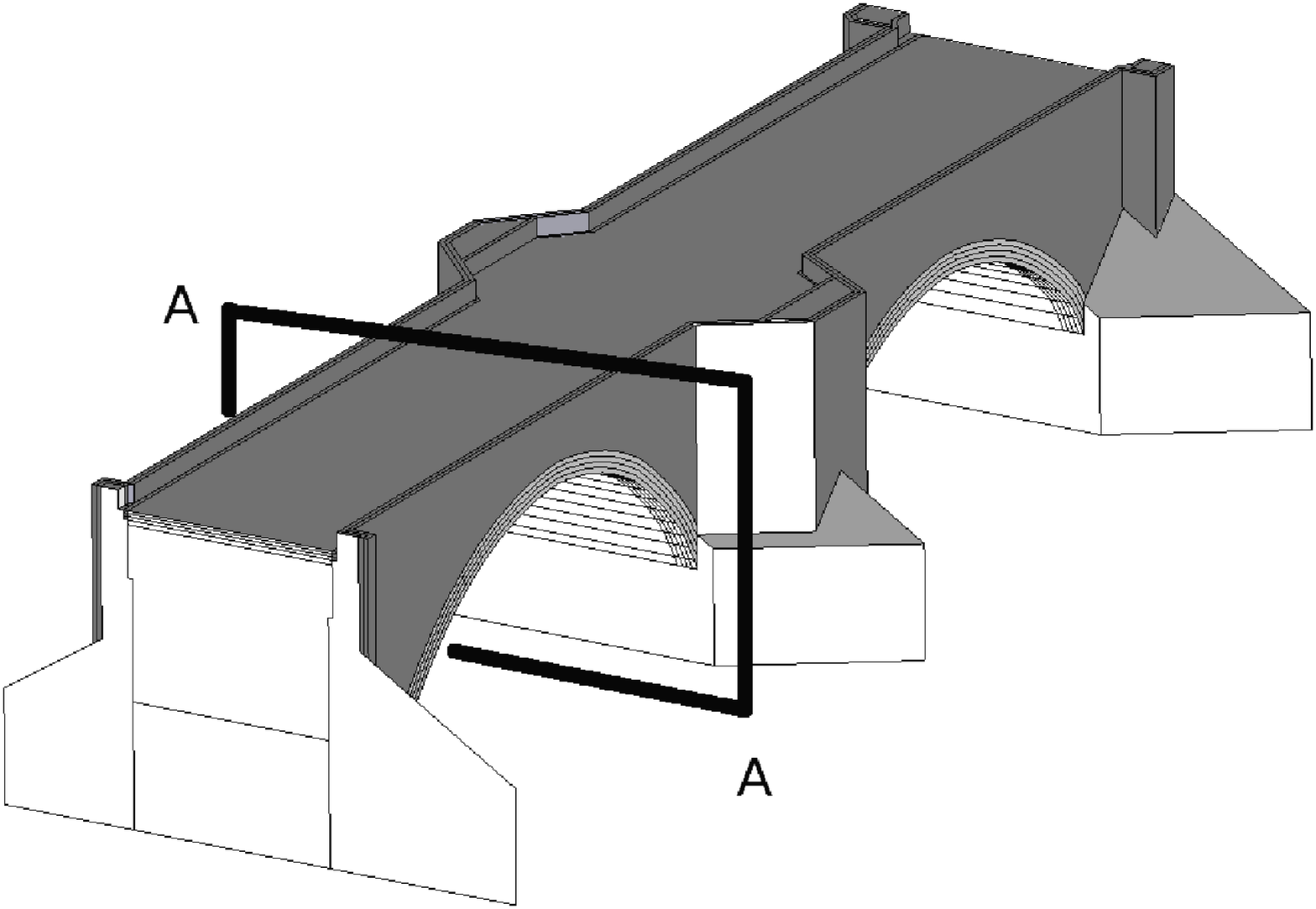}&
\includegraphics[width=65mm,keepaspectratio]{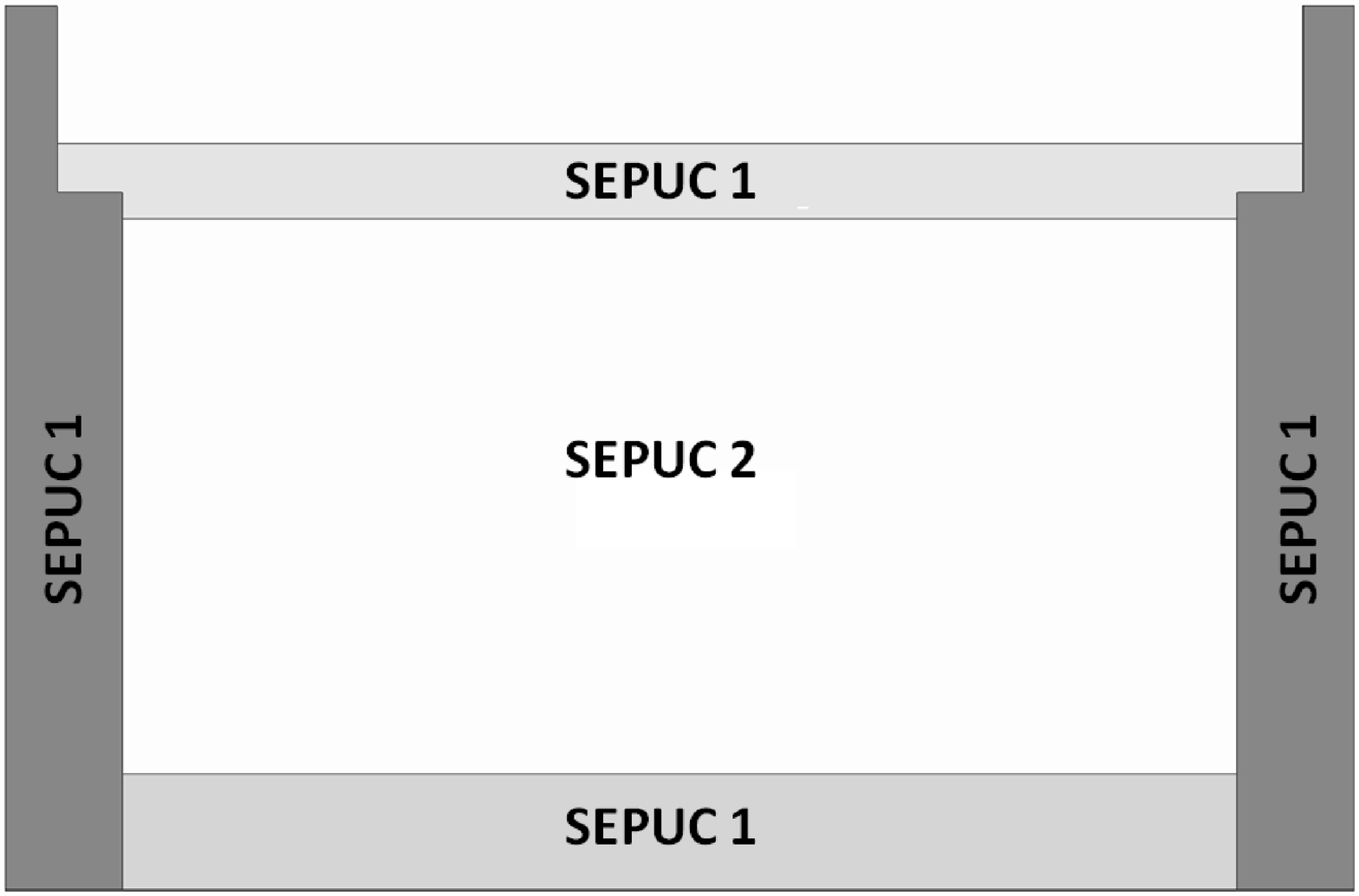}\\
(a) & (b)\\
\end{tabular}
\begin{tabular}{c@{\hspace{15mm}}c}
\multirow{2}{*}[22mm]{\includegraphics[width=80mm,keepaspectratio]{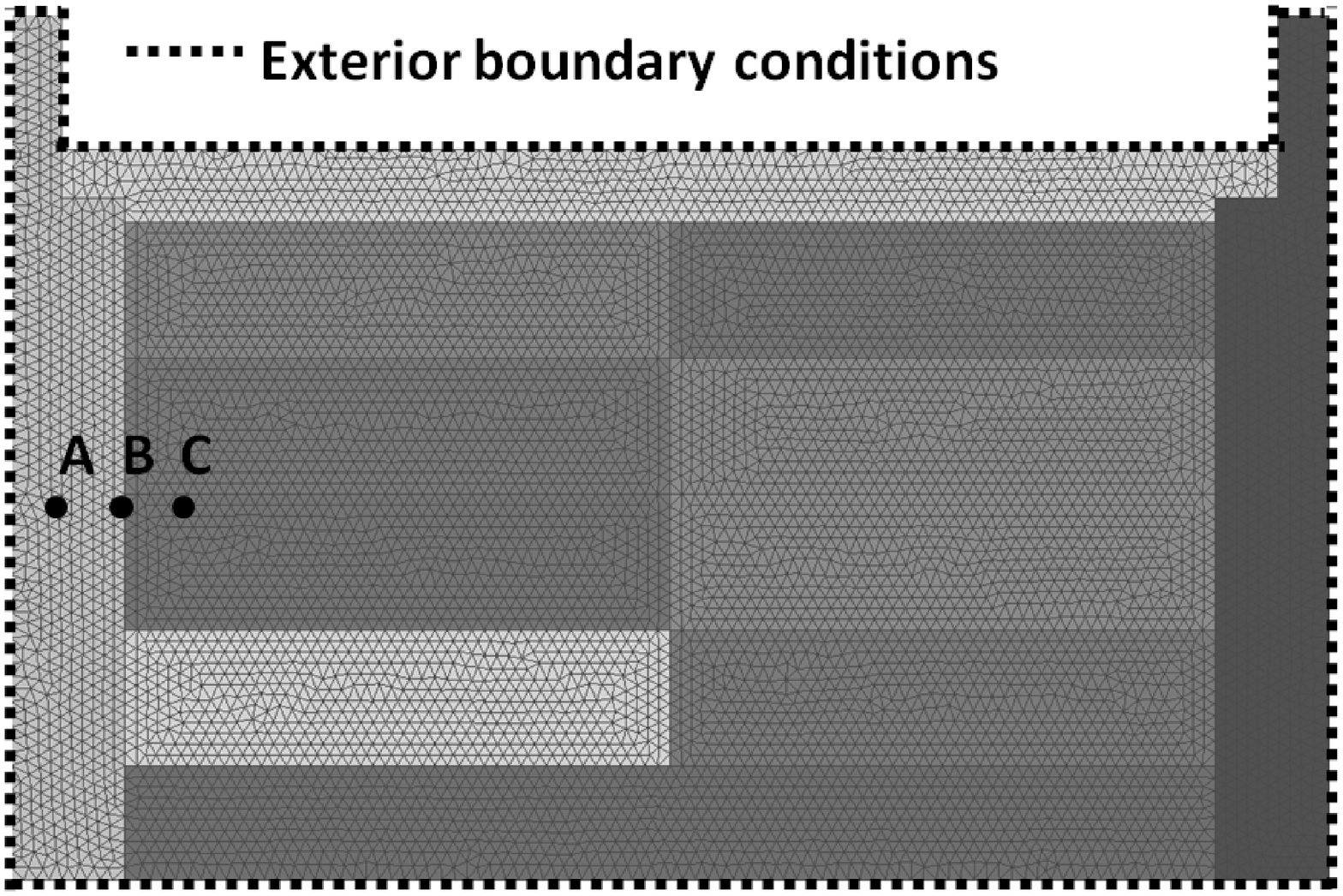}}
& \includegraphics[width=25mm,keepaspectratio]{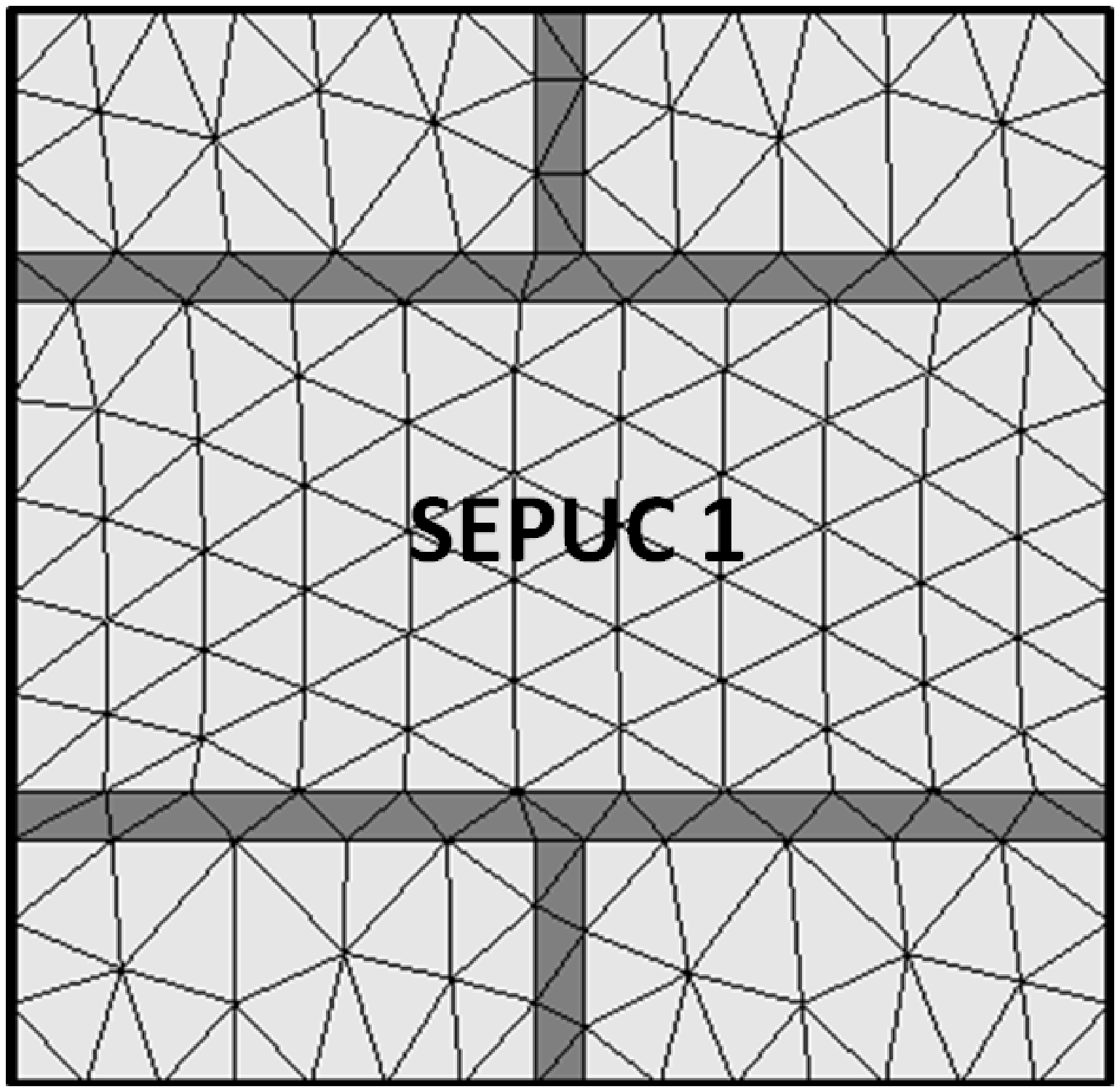}\\
& (d)\\
& \includegraphics[width=25mm,keepaspectratio]{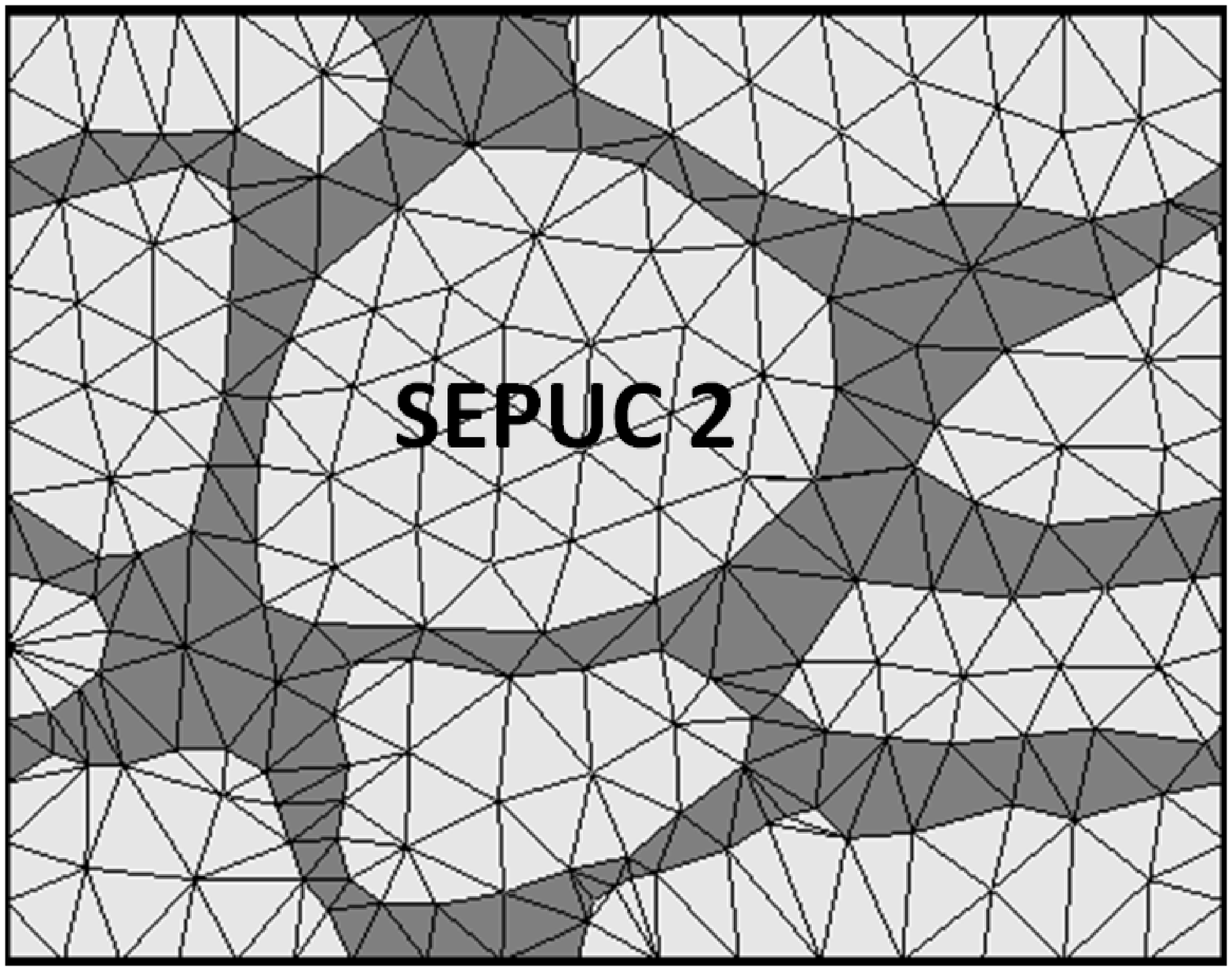}\\
(c) & (e)\\
\end{tabular}
\caption{
(a) Three-dimensional view of Charles Bridge with a two-dimensional A-A section analyzed,
(b) Analyzed section showing material regions with assigned meso-scale unit cells,
(c) Macrostructural mesh with identified loading conditions and decomposition into sub-domains representing individual slave processors ($l_{x}=10.40\,\mathrm{[m]}$, $l_{y}=6.82\,\mathrm{[m]}$),
(d) Mesostructural mesh of regular bonding of masonry (SEPUC 1: $l_{x}=0.45\,\mathrm{[m]}$, $l_{y}=0.44\,\mathrm{[m]}$),
(e) Mesostructural mesh of irregular quarry masonry (SEPUC 2: $l_{x}=0.45\,\mathrm{[m]}$, $l_{y}=0.35\,\mathrm{[m]}$).
}
\label{fig:decomp}
\end{center}
\end{figure}

Fig.~\ref{fig:decomp}(a) displays a three-dimensional segment of
Charles Bridge examined in the original three-dimensional static
calculation~\cite{Novak:ES:2007}. A two-dimensional cut through the
mid part of a Charles Bridge arch examined for the parallel
computation appears in Fig.~\ref{fig:decomp}(b) together with four
material regions. These are associated with two heterogeneity systems
of the bridge, one representing a regular sand stone masonry of side
walls, fence and arches and the other corresponding to an irregular
quarry masonry made of arenaceous marl blocks and sand and black lime
mortar filling the inner part of the bridge. For simplicity, the
bridge deck was assigned the regular pattern. The corresponding
periodic unit cells employed for the meso-scale analysis are plotted
in Figs.~\ref{fig:decomp}(d) and \ref{fig:decomp}(e), respectively.

The finite element mesh used at the macro-level is evident from
Fig.~\ref{fig:decomp}(c) featuring $7,081$ nodes and $13,794$
triangular elements with a single integration point thus amounting to
the solution of $13,794$ meso-problems at each macroscopic time
step. This figure also shows decomposition of the macro-problem into
$12$ slave processors. The numbers of elements in individual
sub-domains being equal to the number of meso-problems handled by the
assigned slave processor are listed in Table \ref{tabdecomp}.  It
should be noted that the assumed decomposition of the macro-problem is
not ideal. In comparison with domain decomposition methods, the
macro-problem has to be split with respect to the heterogeneity of the
material resulting in the variation of number of elements in
individual sub-domains between $1046$ and $1748$.

\begin{table}
\begin{center}
\begin{tabular}{lllllllll}
\hline
Processor No. & 1 & 2 & 3 & 4 & 5 & 6 & 7 & 8\\
& 9 & 10 & 11 & 12 & - & - & - & -\\
\hline
No. of & 1218 & 1748 & 1046 & 1052 & 1214 & 1210 & 1052 & 1054 \\
meso-problems & 1046 & 1052 & 1054 & 1048 & - & - & - & -\\
\hline
\end{tabular}
\caption{Decomposition of the macro-problem into sub-domains.}
\label{tabdecomp}
\end{center}
\end{table}

The number of elements in the two meso-problems amounts to $265$
($160$ nodes) for SEPUC 1 and to $414$ ($239$ nodes) for SEPUC 2,
respectively. Similarly to the macro-problem, the meso-problems have
to account for the material heterogeneity. Clearly, the ideal speedup
and load balancing are obtained when the decomposition of the
macro-problem reflects the meso-problem meshes. However, this is
considerably more difficult when compared to the classical mesh
decomposition.

The actual analysis was performed on a cluster built at our
department. Each node of the cluster is a single processor personal
computer Dell Optiplex GX$620$ equipped with $3.54$ GB of RAM. The
processors are Intel Pentium with the frequency $3.4$ GHz. The cluster
is based on Debian linux $5.0$ and $32$-bit architecture.

\begin{figure} [ht!]
\begin{center}
\begin{tabular}{c@{\hspace{10mm}}c}
\includegraphics*[width=60mm,keepaspectratio]{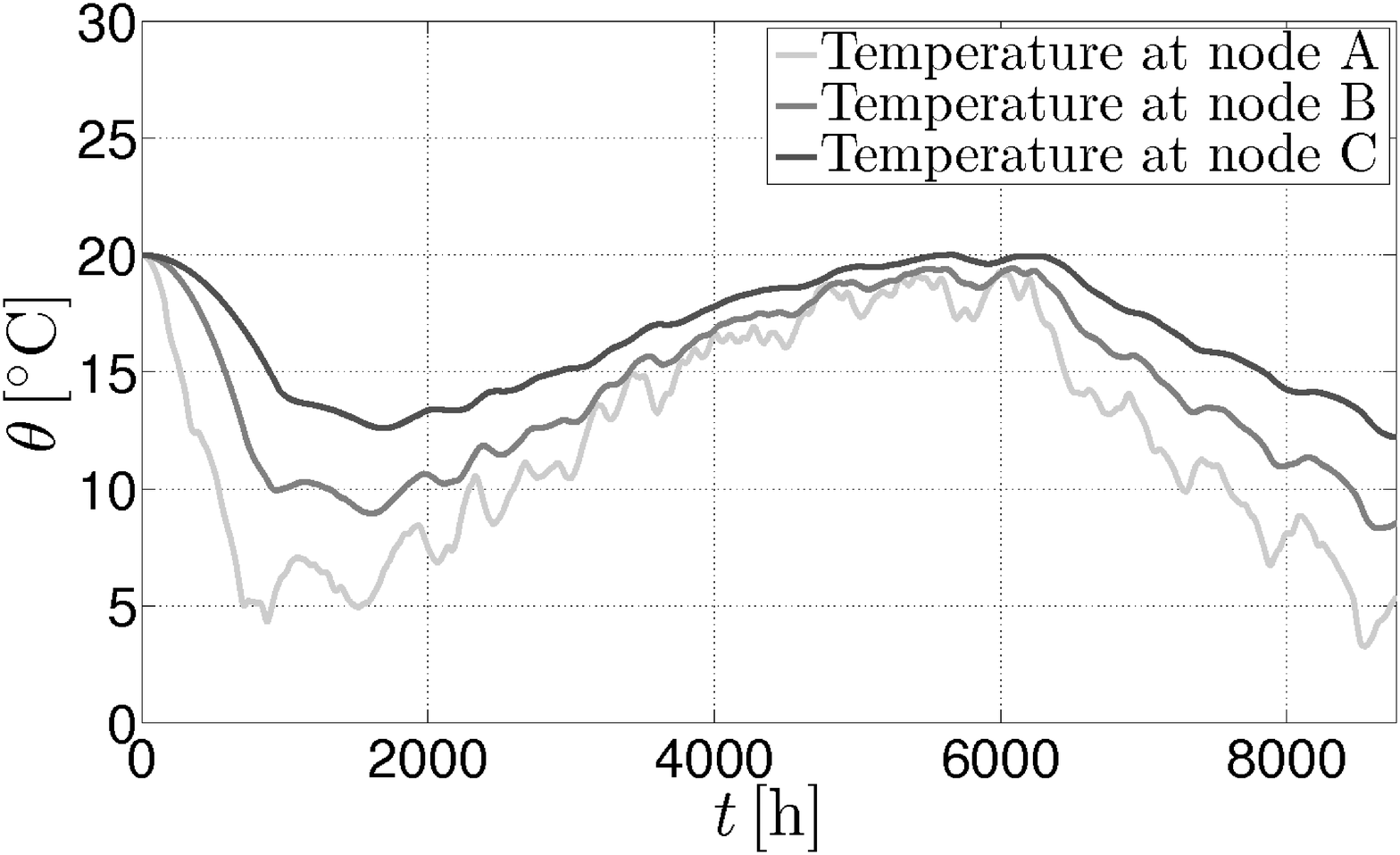}&
\includegraphics*[width=60mm,keepaspectratio]{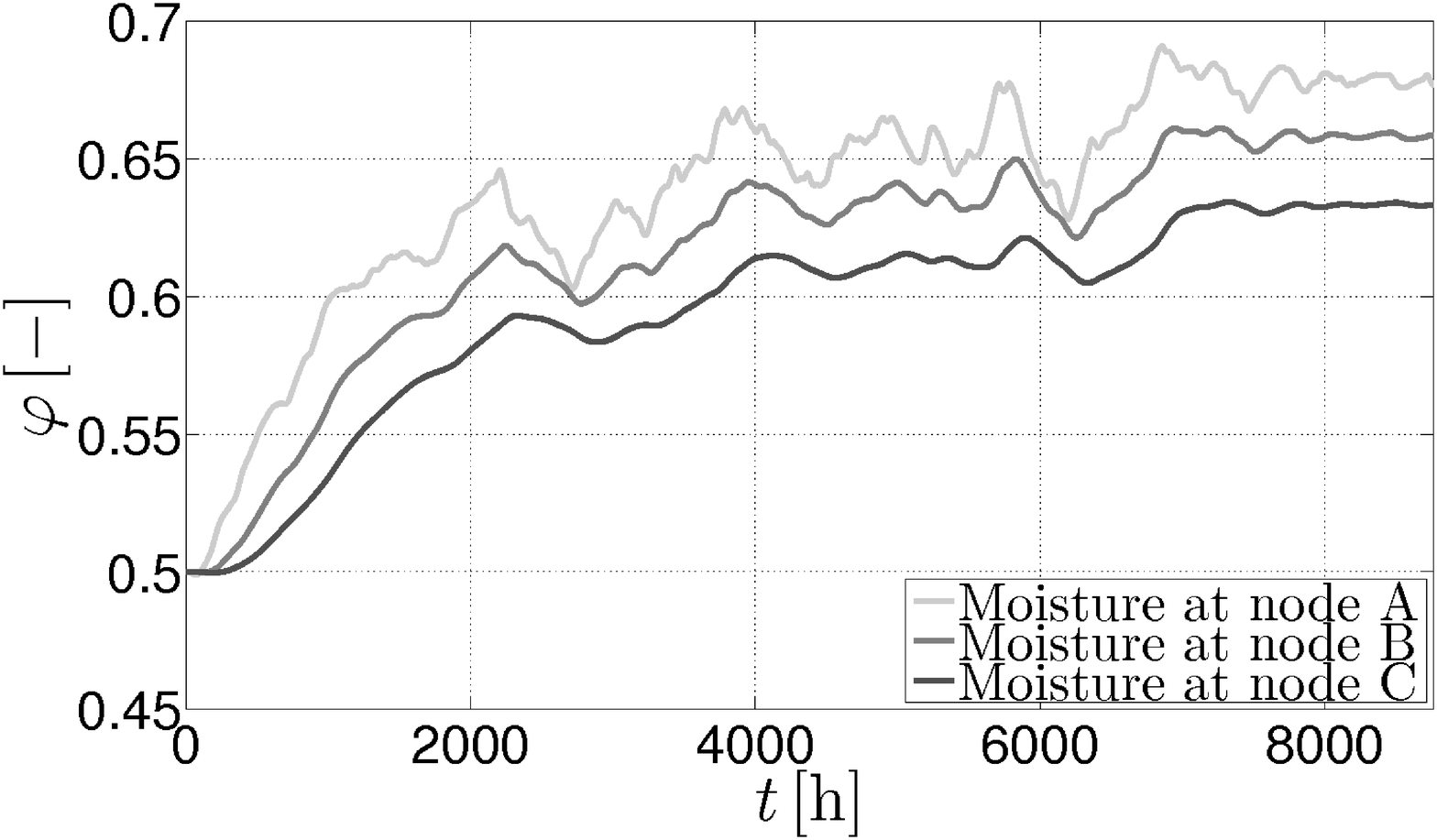}\\
(a)&(b)
\end{tabular}
\end{center}
\caption{Evolution of macroscopic (a) temperature, (b) moisture} \label{fig:evol}
\end{figure}

Although various mesoscopic heterogeneity patterns were properly
accounted for, the material data of individual constituents (bricks or
stones and mortar) were taken the same for both the outer and inner
part of the bridge, recall Table~\ref{tab:matpar} for specific values.
The initial conditions on the macro-scale were set equal to
$\varphi=0.5$ [-] and $\theta=20$ [$^\circ$C] and the year round
variation of moisture and temperature in Fig.~\ref{fig:BC} was imposed
onto all outer surfaces of the bridge, see
Fig.~\ref{fig:decomp}(c). Owing to the computational demands the
macroscopic time increment was set equal to $10$ [h], which agreed
with the real computational time equal to $2.08$ minute for each time
step. In view of the results presented in Section~\ref{subsec:InfFEM},
recall Fig.~\ref{fig:Comp}, this justifies, although at the loss of
accuracy at the initial stage of computation, the use of full
multi-scale scheme. One particular example of the resulting evolutions
of macroscopic temperature and moisture at the selected nodes labeled
in Fig.~\ref{fig:decomp}(c) is seen in Fig.~\ref{fig:evol}.

\section{Conclusions}\label{sec:Conclusions}
A fully coupled multi-scale analysis of simultaneous heat and moisture
transport in masonry structures was implemented in the framework of
FE$^2$ computational strategy. Two particular issues were addressed:
the influence of the finite size of SEPUC when running the transient
analysis on both scales and the way of introducing loading on the
macro-scale. While the former one plays a significant role in the
estimates of macroscopic response, the latter one proves important
only in the initial stages of loading.

Special attention was accorded to the implementation of FE$^2$ concept
in the parallel format employing the master-slave approach. Although
not qualitatively fully acceptable, the two-dimensional example of
Charles Bridge raised a number of questions to the solution efficiency
particularly with reference to a proper subdivision of the analyzed
macro-domain and local finite element mesh of individual meso-scale
SEPUCs. The present findings summarized in
Section~\ref{subsec:Parallelization} will be utilized in a fully
three-dimensional analysis being the subject of our current research
effort.

\section*{Acknowledgment}
This outcome has been achieved with the financial support of the Czech
Science Foundation, project No. 105/11/0411, by the Ministry of
Industry and Trade of the Czech Republic though FR-TI1/381 project,
and partially also by the research project CEZ~MSM~6840770003.

\appendix\section{}\label{app:A}
The list of material parameters to be obtained experimentally are
stored in Table~\ref{tab:matpar}. The transport coefficients that
enter Eqs.~\eqref{Kun01}~and~\eqref{Kun02} are provided by
\begin{itemize}
\item $w$ - water content $\mathrm{[kgm^{-3}]}$,
    \begin{equation}
    w = w_{f}\frac{(b-1)\varphi}{b-\varphi}, \label{eq:DTC1}
    \end{equation}
    where $w_{f}$ is the free water saturation and $b$ is the
    approximation factor, which must always be greater than one. It
    can be determined from the equilibrium water content ($w_{80}$) at
    $0.8$ [-] relative humidity by substituting the corresponding
    numerical values in equation~\eqref{DTC1}.
\item $\delta_{p}$ - water vapor permeability
    $\mathrm{[kgm^{-1}s^{-1}Pa^{-1}]}$,
    \begin{equation}
    \delta_{p} = \frac{\delta}{\mu},
    \end{equation}
    where $\mu$ is the water vapor diffusion resistance factor and $\delta$ is the vapor diffusion
    coefficient in air $\mathrm{[kgm^{-1}s^{-1}Pa^{-1}]}$ given by
    \begin{equation}
    \delta = \frac{2.306\cdot{}10^{-5 }\,p_{a}}{R_{v}\,(\theta+273.15)\,p}\left(\frac{\theta+273.15}{273.15}\right )^{1.81},
    \end{equation}
    with $p$ set equal to atmospheric pressure $p_{a}=101325$ [Pa] and
    $R_{v}$ = $R$/$M_{w} = 461.5$ [$\mathrm{Jkg^{-1}K^{-1}}$]; $R$
    is the gas constant ($8314.41$ [Jmol$^{-1}$K$^{-1}$]) and $M_w$
    is the molar mass of water ($18.01528$ [kgmol$^{-1}$]).
\item $D_{\varphi}$ - liquid conduction coefficient
    $\mathrm{[kgm^{-1}s^{-1}]}$,
    \begin{equation}
    D_{\varphi}=D_{w}\frac{\mathrm{d}w}{\mathrm{d}\varphi},
    \end{equation}
    where $D_{w}$ is the capillary transport coefficient given by
    \begin{equation}
    D_{w}=3.8\left(\frac{A}{w_{f}}\right
    )^{2}\cdot{}10^{3w/(w_{f}-1)},
    \end{equation}
    where the derivative of the moisture storage function $\displaystyle{\frac{\mathrm{d}w}{\mathrm{d}\varphi}}$
    is obtained by differentiating Eq.~\eqref{DTC1}.
\item $\lambda$ - thermal conductivity $\mathrm{[Wm^{-1}K^{-1}]}$,
    \begin{equation}
    \lambda=\lambda_{0}\left(1+\frac{b_{\mathrm{tcs}}w}{\rho_{s}}\right),
    \end{equation}
    where $\lambda_{0}$ is the thermal conductivity of dry building
    material, $\rho_{s}$ is the bulk density and $b_{\mathrm{tcs}}$ is
    the thermal conductivity supplement.
\item $p_{\mathrm{sat}}$ - water vapor saturation pressure
    $\mathrm{[Pa]}$,
    \begin{equation}
    p_{\mathrm{sat}}=611\,\exp\left(\frac{a\,\theta}{\theta_{0}+\theta}\right),
    \end{equation}
    where
    \begin{equation}
    \begin{array}{ccc}
    a=22.44 & \theta_{0}=272.44\,[^{\circ}\mathrm{C}] &
    \theta<0\,[^{\circ}\mathrm{C}]\\[0.5cm]
    a=17.08 & \theta_{0}=234.18\,[^{\circ}\mathrm{C}] &
    \theta\geq0\,[^{\circ}\mathrm{C}]
    \end{array}
    \end{equation}
\item $h_v$ - evaporation enthalpy of water $\mathrm{[Jkg^{-1}]}$
    \begin{equation}
    h_v = 2.5008 \cdot{} 10^{6} \left(\frac{273.15}{\theta}\right)^{(0.167 +
    3.67 \cdot{} 10^{-4}\theta )}.
    \end{equation}
\end{itemize}

\bibliographystyle{elsarticle-num}
\bibliography{liter_arxiv}

\end{document}

%% file: format.tex
\newcommand{\eqref}[1]{(\ref{eq:#1})}

\newcommand{\bmath}[1]{\mbox{\boldmath$#1$}}
\newcommand{\vek}[1]{\bmath{#1}}

\newcommand{\mat}[1]{\mathchoice{\displaystyle\mathbf#1}
{\textstyle\mathbf#1}{\scriptstyle\mathbf#1}
{\scriptscriptstyle\mathbf#1}}

\newcommand{\tensf}[1]{\bmath{\mathsf{#1}}}          